\documentclass[reprint,superscriptaddress,showpacs,amsmath,amssymb,aps,prl]{revtex4-1}
\usepackage{graphicx}
\usepackage{dcolumn}
\usepackage{bm}
\usepackage{hyperref}
\hypersetup{colorlinks=true, citecolor=blue, urlcolor=blue, linkcolor=blue}
\usepackage{amssymb}
\usepackage{color}
\usepackage{graphicx}
\usepackage{amsmath}
\usepackage{mathrsfs}
\usepackage{times}
\usepackage{subeqnarray}
\usepackage{cases}
\usepackage{bm}
\usepackage{diagbox}
\usepackage{booktabs}
\usepackage[table,xcdraw]{xcolor}
\usepackage{colortbl}
\usepackage{epstopdf}
\definecolor{tabcolor}{rgb}{.105,.410,.113}
\usepackage{overpic}
\usepackage{algorithm}
\usepackage{algpseudocode} 

\begin{document}
\title{Catalytic evolution of cooperation in a population with behavioural bimodality}
	
\author{Anhui Sheng}
\affiliation{School of Physics and Information Technology, Shaanxi Normal University, Xi'an 710061, P. R. China}
\author{Jing Zhang}
\affiliation{College of Information Science and Technology, Donghua University, Shanghai 201620, P. R. China}
\affiliation{School of Physics and Information Technology, Shaanxi Normal University, Xi'an 710061, P. R. China}
\author{Guozhong Zheng}
\affiliation{School of Physics and Information Technology, Shaanxi Normal University, Xi'an 710061, P. R. China}
\author{Jiqiang Zhang}
\affiliation{School of Physics, Ningxia University, Yinchuan 750021, P. R. China}
\author{Weiran Cai}
\affiliation{School of Computer Science, Soochow University, Suzhou 215006, P. R. China}
\author{Li Chen}
\email[Email address: ]{chenl@snnu.edu.cn}
\affiliation{School of Physics and Information Technology, Shaanxi Normal University, Xi'an 710061, P. R. China}

\begin{abstract}
The remarkable adaptability of humans in response to complex environments is often demonstrated by the context-dependent adoption of different behavioral modes. However, the existing game-theoretic studies mostly focus on the single-mode assumption, and the impact of this behavioral multimodality on the evolution of cooperation remains largely unknown. Here, we study how cooperation evolves in a population with two behavioral modes. Specifically, we incorporate Q-learning and Tit-for-Tat (TFT) rules into our toy model, where prisoner's dilemma game is played and we investigate the impact of the mode mixture on the evolution of cooperation. While players in Q-learning mode aim to maximize their accumulated payoffs, players within TFT mode repeat what their neighbors have done to them. In a structured mixing implementation where the updating rule is fixed for each individual, we find that the mode mixture greatly promotes the overall cooperation prevalence. The promotion is even more significant in the probabilistic mixing, where players randomly select one of the two rules at each step. Finally, this promotion is robust when players are allowed to adaptively choose the two modes by real-time comparison. In all three scenarios,  players within the Q-learning mode act as catalyzer that turns the TFT players to be more cooperative, and as a result drive the whole population to be highly cooperative. The analysis of Q-tables explains the underlying mechanism of cooperation promotion, which captures the ``psychologic evolution" in the players' mind.   Our study indicates that the variety of behavioral modes is non-negligible, and could be crucial to clarify the emergence of cooperation in the real world.
\end{abstract}
	
\date{\today }
\maketitle

\section{1. Introduction}\label{sec:introduction}
	
Cooperation exists in a wide range of natural and human societies~\cite{smith1997major}, from intra- and inter-species cooperation in ecology to the labor division of individuals in socio-economic activities~\cite{Zelenski2015CooperationII,Cheney2011Extent,dawkins2016selfish}. While the cooperation is widely observed in human beings and other animals, it's actually counterintuitive since defection in many scenarios is the rational choice in terms of higher payoff, especially in some social dilemma situations~\cite{rapoport1965prisoner}. The key question is: how does cooperation evolve? which is also listed as one of the twenty-five grand questions in this century~\cite{Elizabeth2005How}. Understanding the underlying mechanism of cooperation is of fundamental importance in theory and has practical implications for solving many pressing challenges, such as the climate negotiation, the trade war, and regional conflicts, etc~\cite{Garrett1968The,Kollock1998Social,Manfred2006Stabilizing}. 

Within the framework of evolutionary game theory, plenty of works based on the prototypical model such as prisoner's dilemma game (PDG)~\cite{poundstone1993prisoner,doebeli2005models,axelrod1980effective,perc2008social,axelrod1980more,milinski1998working} have been done to address the mechanisms for the emergence of cooperation and its maintenance. Important progresses have been made, where several mechanisms are revealed~\cite{Nowak2006Five,Perc2017statistical}, such as kin selection~\cite{Hamilton1964the}, direct~\cite{trivers1971evolution} and indirect reciprocity~\cite{nowak1998evolution}, network reciprocity~\cite{nowak1992evolutionary, Wang2013Interdependent}, dynamical reciprocity~\cite{Liang2022dynamical}, group selection~\cite{keller1999levels,smith1964group}, punishment and reward~\cite{Sigmund2001Reward,Xia2023Reputation}, and social diversity~\cite{perc2008social,Santos2008Social,Liang2021Social} etc.

Recently, the reinforcement learning (RL)~\cite{Watkins1992Technical,Sutton2018reinforcement} as a new paradigm has been applied to the game-theoretic studies, which has garnered increasing attention. Distinct from the previous framework of social learning~\cite{Bandura1977social}, where individuals imitate the strategies of those who have higher payoffs, the decision-making in RL is based upon past experience to maximize cumulative rewards through interaction with the environment~\cite{Sutton2018reinforcement, silver2018general}. The rationale of RL is deeply rooted within human cognition and has solid experimental evidence from neuroscience~\cite{Lee2012neural, Rangel2008A}. New insights have been obtained by RL  for the emergence of cooperation~\cite{Zhang2020Oscillatory,Wang2022Levy,Wang2023Synergistic,He2022migration,Ding2023Emergence,Geng2022Reinforcement,Zhang2024emergence} and trust~\cite{zheng2023decoding}, resource allocation~\cite{Andrecut2001q, Zhang2019reinforcement, zheng2023optimal, Zhang2024self}, and other collective human behaviors~\cite{Zhang2020Oscillatory,Tomov2021multi,Shi2022analysis,Ding2024Emergence}. 
Till now, Q-learning~\cite{Watkins1989learning,Watkins1992Technical} is the commonly adopted RL algorithm ~\cite{watkins1992q,van2016deep,wunder2010classes,Shi2022analysis}, where there is a Q-table scoring different actions within different states, which provides the explainability of the algorithm. By monitoring the evolution of Q-table, we are able to study the preference change, which captures the psychological evolution in the players' mind. This is critical for understanding the logic of human behaviors, and is lacking in the traditional framework. 

Although huge progresses are made in the paradigm of both social learning and reinforcement learning, most of the game-theoretic work assumes single-mode scenario. Namely, individuals make their decision according to a single behavior mode, e.g. the imitation rule~\cite{Szabo2007evolutionary, Perc2017statistical} (e.g. Fermi updating rule, Moran rule etc.), the reinforcement learning~\cite{Sutton2018reinforcement} (Q-learning, SASA, the Bush-Mosteller model etc.), and reactive strategies~\cite{axelrod1981evolution,nowak1992tit,nowak1993strategy} (e.g. tit-for-tat and generous tit-for-tat). This assumption is, however, in contrast to observations in the real world, where individuals behave within one mode in a given scenario, but switch to a different mode once the scenario is changed. Behavioral experiments~\cite{Traulsen2010Human} indicates that we human actually adopt multiple behavioral modes and the proportions of different modes are also time-varying. In fact, such behavioral multimodality is crucial to the adaptability to complex surroundings, and is the key to our survival.
	
Some recent game-theoretic work has started to account for the diversity of behavioral modes, in order to understand their impact on the emergence of cooperation. Previous studies show the cooperation prevalence could either be promoted by mixed mode of payoff-based imitation and conformity-based rule~\cite{Szolnoki2015Conformity, Szolnoki2018Competition}, or be deteriorated by mode mixture of imitation and innovation~\cite{Amaral2018Heterogeneous}. Interestingly, when the population is composed of normal players and zealots~\cite{Masuda2012evolution} or ``good samaritans"~\cite{Zheng2022probabilistic}, this mixture could unexpectedly lift the prevalence of cooperation or fairness. In Ref.~\cite{ma2023emergence}, Ma et al. study the evolution of cooperation within bimodality of Fermi rule and TFT, and they show the cooperation level is significantly promoted in both quenched and annealed manners, and this promotion is robust to the underlying topology. Han et. al propose a hybrid learning combining social learning and the Bush-Mosteller model as self-learning~\cite{Han2022hybrid}, they reveal cooperation prevalence is promoted within some parameter regions. As the new paradigm of RL gradually releases its potential in explaining cooperation, we are interested in the following questions: what is impact of the behavioral multimodality on the evolution of cooperation when the reinforcement learning is engaged?
	
In this work, we focus on the impact of behavioral multimodality in the evolution of cooperation. Specifically, we investigate the bimodal mixture of two fundamentally different update rules --- TFT and Q-learning. As a new paradigm, players within Q-learning mode resort to continuously policy modification as introspection, while players within TFT mode just repeat what their opponents have done to them~\cite{nowak1993strategy}. The mode mixture is implemented in three manners: structural mixing, probabilistic mixing, and adaptive mixing. In all three implementations, the prevalence of cooperation is significantly improved compared to either single-mode scenario. By monitoring the Q-table, we clarify the mechanism of the cooperation promotion and the impact of the bimodality, which provides valuable insights into the psychologic evolution of population.
 
The rest of this paper is structured as follows. Sec. 2 describe our model with three implementations. Sec. 3 shows the results of the cooperation evolution. Sec. 4 discusses the underlying mechanism behind the cooperation promotion. Finally, Sec. 5 provides conclusions and discussions of this work.

\begin{table}[]
		\begin{tabular}{c|cc}
			\arrayrulecolor{tabcolor}\toprule [1.4pt]
			\hline
			\diagbox{State}{Action}& C ($a_{1}$) & D ($a_{2}$) \\
			\midrule [0.5pt]
			\hline
			$s_{1} = (a_{1},0)$  & $Q_{s_{1},a_{1}}$& $Q_{s_{1},a_{2}}$\\
                         $s_{2} = (a_{2},0)$  & $Q_{s_{2},a_{1}}$& $Q_{s_{2},a_{2}}$\\
			$s_{3} = (a_{1},1)$  & $Q_{s_{3},a_{1}}$& $Q_{s_{3},a_{2}}$\\
			$s_{4} = (a_{2},1)$  & $Q_{s_{4},a_{1}}$& $Q_{s_{4},a_{2}}$\\
			$s_{5} = (a_{1},2)$  & $Q_{s_{5},a_{1}}$& $Q_{s_{5},a_{2}}$\\
			$s_{6} = (a_{2},2)$  & $Q_{s_{6},a_{1}}$& $Q_{s_{6},a_{2}}$\\
			$s_{7} = (a_{1},3)$  & $Q_{s_{7},a_{1}}$& $Q_{s_{7},a_{2}}$\\
			$s_{8} = (a_{2},3)$  & $Q_{s_{8},a_{1}}$& $Q_{s_{8},a_{2}}$\\
			$s_{9} = (a_{1},4)$  & $Q_{s_{9},a_{1}}$& $Q_{s_{9},a_{2}}$\\
			$s_{10} = (a_{2},4)$& $Q_{s_{10},a_{1}}$&$Q_{s_{10},a_{2}}$\\ 
			\hline
			\bottomrule[1.4pt]
		\end{tabular}
		\caption{Q-table for every player within the Q-learning mode. The state of a given player $s_{1,...,10}$ corresponds to the 10 combination of player's own action ($a_1$ or $a_2$)  and the number of cooperators in his four nearest neighbours.}
		\label{tab:Qtable}
\end{table}

\section{2. Methods}\label{sec:model}

We consider a population of $N$ individuals who play the Prisoner's Dilemma Game (PDG) on a square lattice $L\times L=N$ with periodic boundary conditions, and $L=100$ if not stated otherwise. For each paired players, each can choose to be either cooperator (C) or defector (D), both players get a reward $R$ for mutual cooperation, mutual defection yields the punishment $P$ to each, and mixed encounter gives the cooperator the sucker's payoff $S$ yet the temptation $T$ for the defector. The order of these four payoffs $T>R>P>S$ and $2R>T+S$ are required for the PDG. For clarity, the payoff matrix of the PDG is as follows:
\begin{equation}
		\Pi=\left(\begin{array}{ll}
			\Pi_{C C} & \Pi_{C D} \\
			\Pi_{D C} & \Pi_{D D}
		\end{array}\right)=\left(\begin{array}{cc}
			R & S \\
			T & P
		\end{array}\right),
\label{eq:payoff}
\end{equation}
where the values of these four payoffs are reconfigured as $R=1$, $S=-r$, $T=1+r$, $P=0$ in our study, making a strong version of PDG. The dilemma lies in the fact that although mutual cooperation is more beneficial for the collective payoff, they still defect driven by the self-interest and mutual defection is the only Nash equilibrium of the game~\cite{nowak1992evolutionary}. The value of $r$ $(0<r<1.0)$ controls the strength of the social dilemma, the larger value of $r$, the stronger the social dilemma.

\begin{figure*}[htbp]
\centering	
\begin{overpic}[width=0.3\linewidth]{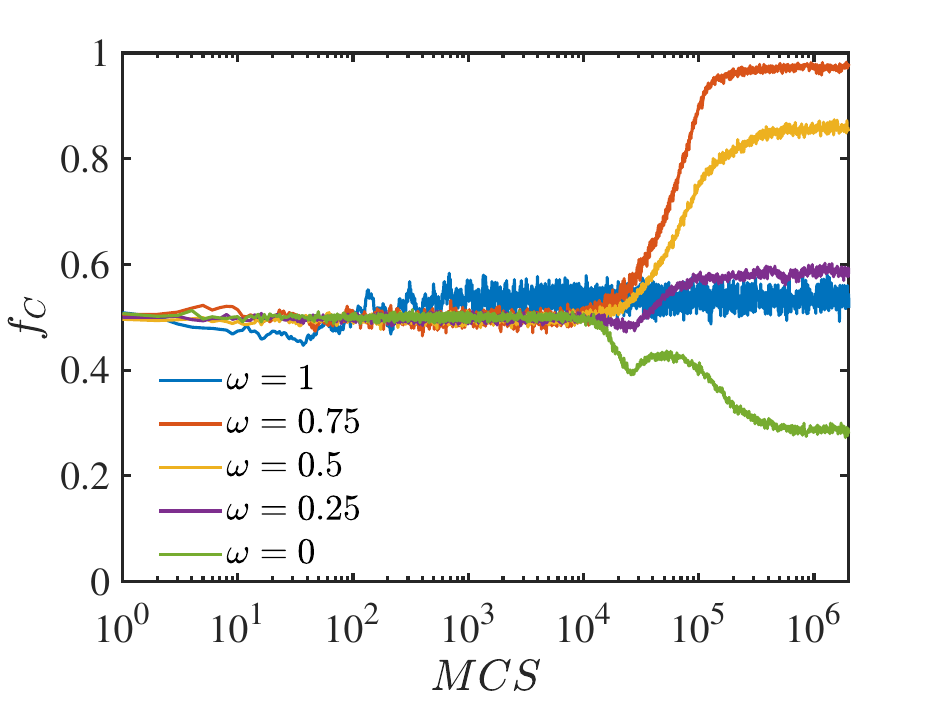}\put(2.5,65){(a)}\end{overpic}
\begin{overpic}[width=0.3\linewidth]{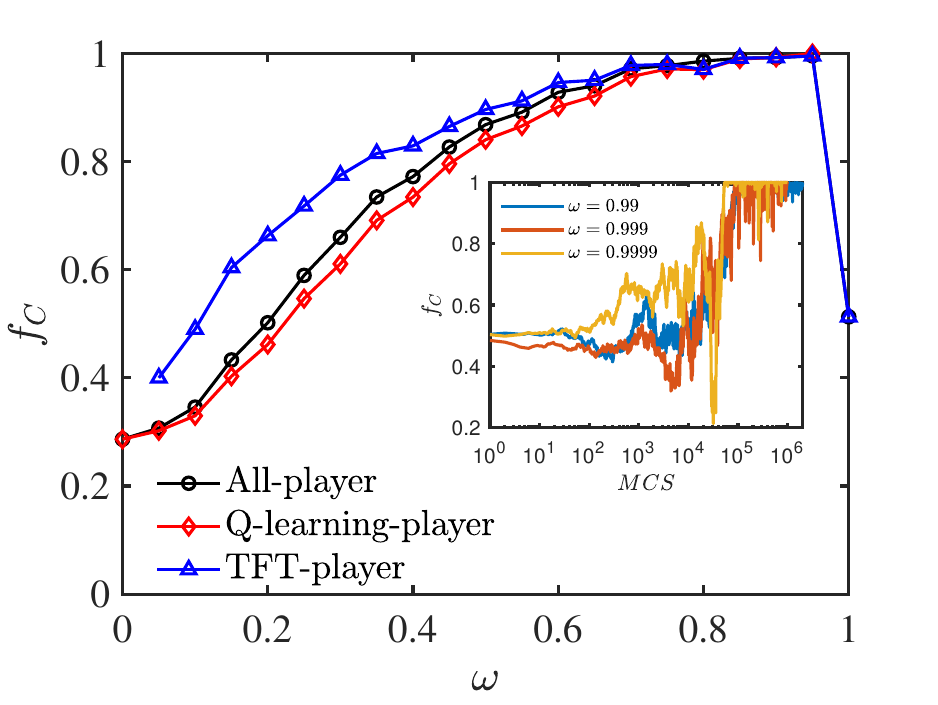}\put(2.5,65){(b)}\end{overpic}
\begin{overpic}[width=0.3\linewidth]{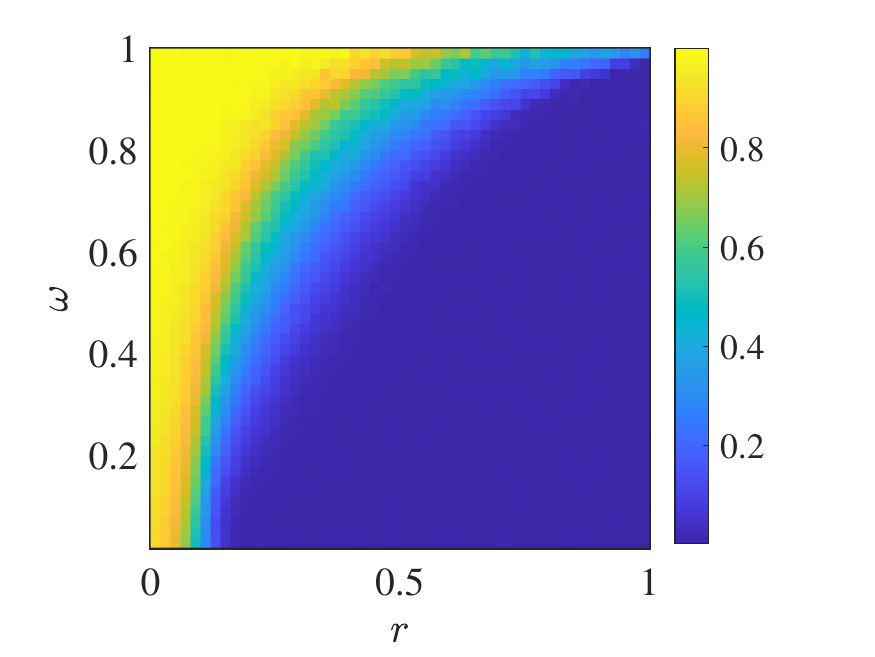}\put(2.5,65){(c)}\end{overpic}
\caption{The evolution of cooperation with SM. $\omega$ is the proportion of players within the TFT mode.
	(a) Time series of cooperation level $f_C$ for different proportions of players within the TFT mode.
	(b) The prevalence of cooperation $f_C$ as a function of mode proportion $\omega$, separately computed for people within TFT mode, Q-learning mode, and for the whole population; the inset provides three time series of $f_C$ for the whole population for three cases of $\omega\rightarrow 1$.
	(c) Heat map for the cooperation prevalence $f_C$ in the $r-\omega$ parameter space.
	The game parameter $r=0.1$ in (a) and (b). 
	Other parameters: $\varepsilon = 0.01$, $\alpha =0.1$, and $\gamma =0.9$.
}
\label{fig:SM}
\end{figure*}

In our model, individuals play the game within two possible behavioral modes, i.e. the TFT rule and Q-learning. Within TFT mode, players just repeat what their neighbors play against it. Within the Q-learning mode~\cite{watkins1992q}, each player is equipped with a Q-table that can be regarded as a policy and guides one's decision making, see Table~\ref{tab:Qtable}. The idea behind Q-table is to score the different action $a\in A$ within the different state  $s\in S$, reflected in the value of the corresponding item $Q_{s,a}$. In our model, the action set $A=\{C,D\}$, and the state set $S=\{s_1,...,s_{10}\}$, which includes the number of cooperators in its four nearest neighbors and its own action in the last round. The items in the table $Q_{s,a}$ are the action-value function that estimates the value of action $a$ within the given state $s$; if  $Q_{s,a}>Q_{s,\hat{a}}$, this means that the action $a$ is of higher value than action $\hat{a}$ within state $s$, and is suggested to be chosen by Q-learning. Within Q-learning mode, players revise their Q-tables to maximize their accumulated payoffs.

In our bimodal behaviors model, we consider three mixing implementations -- \emph{structure mixing} (SM), \emph{probabilistic mixing} (PM), and \emph{adaptive mixing} (AM). In the SM scenario,  each player is assigned with one of the two behavioral modes at the very beginning, specifically the TFT mode is chosen with probability $\omega$ and Q-learning with $1-\omega$, and they behave in the same mode throughout the whole evolution. 
Instead, players randomly choose one of the two modes at each round in the PM, with the same probability denotation. Different from these two implementations where the probabilities are fixed, the AM procedure follows an adaptive manner: when the payoff of player $i$ is lower than the value in previous round i.e. $\Pi_i(t)<\Pi_i(t-1)$, this unsatisfactory outcome makes him switch from the adopted mode to the other.

To be specific, we first detail the procedure of SM as follows:

(1) At the beginning ($t=0$), every player chooses the TFT and Q-learning mode at random, respectively with the probability $\omega$ and $1-\omega$. Once the mode is chosen, the player will stick to it throughout the whole evolution. For initialization, each player is endowed an action $a\in A$ with an equal probability, and for those within the Q-learning mode, each has a Q-table with all items $Q_{s,a}$ being initialized from 0 to 1 independently.

(2) At round $t>0$, if player $i$ is within TFT mode, one of its four neighbors is randomly chosen (say player $j$), and then $i$ adopts the action chosen by player $j$ in the last round, i.e. $a_i(t)=a_j(t-1)$.  

(3) Otherwise, if player $i$ is within the Q-learning mode, he makes his moves according to the Q-learning algorithm. With the probability $\epsilon$, player $i$ randomly choose an action $a\in A$ to make the trial-and-error exploration. Otherwise, player $i$ selects the action based upon 
\begin{equation}
		\begin{aligned}
			a_{i}(t) \leftarrow\arg\max _{a}\left\{Q_{s_{i}, a_{i}}(t)\right\}, a_{i} \in {A},
		\end{aligned}
\end{equation}
where $\arg\max _{a}\left\{Q_{s_{i}, a}(t)\right\}$ is the action corresponding to the maximum Q-value in the row of state $s_i$, where $s_i=(a_i(t-1), n_i^C(t-1))$ and $n_i^C(t-1)$ is the number of cooperators within player $i$'s four neighbors at the end of round $t-1$.

(4) After all players make their moves [i.e. complete step (2) or (3)], those who are within the Q-learning mode proceed to revise their Q-tables as follows:
\begin{equation}
		\begin{aligned}
			Q_{s_{i}, a_{i}}(t\!+\!1)\!=\!(1\!-\!\alpha) Q_{s_{i}, a_{i}}(t)\!+\!\alpha\left[\Pi_{i}(t)\!+\!\gamma \max_{a_{i}^{\prime}} Q_{s_i^{\prime}, a_{i}^{\prime}}(t)\right],
		\end{aligned}
\label{eq:Q-learning}
\end{equation} 
where $s_{i},a_{i} $ represent the current state and action of player $i$, and $s_i^{\prime}=(a_i(t), n_i^C(t))$ is the new state and $a_{i}^{\prime}$ is the new possible action at $t+1$. $\alpha\in (0,1]$ is the learning rate, which reflects the intensity of memory effect; the larger value of $\alpha$, the more amount of Q-value is renewed, meaning more forgetful of the player. 
$\Pi_i(t)$ is the player $i$'s total payoff by playing against its four neighbors at the current round $t$, which follows the payoff matrix Eq. (\ref{eq:payoff}). 
$\gamma \in [0,1]$ is the discount factor, determining the weight of the future reward since $\max_{a_{i}^{\prime}} Q_{s_i^{\prime}, a_{i}^{\prime}}(t)$ is the maximum Q-value one can expect within its new state $s_i^{\prime}$. This completes the Q-table and state update, and one single round is done when there are players within the Q-learning mode. 

(5) Repeat steps(2)-(4) until the system reaches statistically stable state.

\begin{figure*}[htbp]
\centering
\begin{overpic}[width=0.3\linewidth]{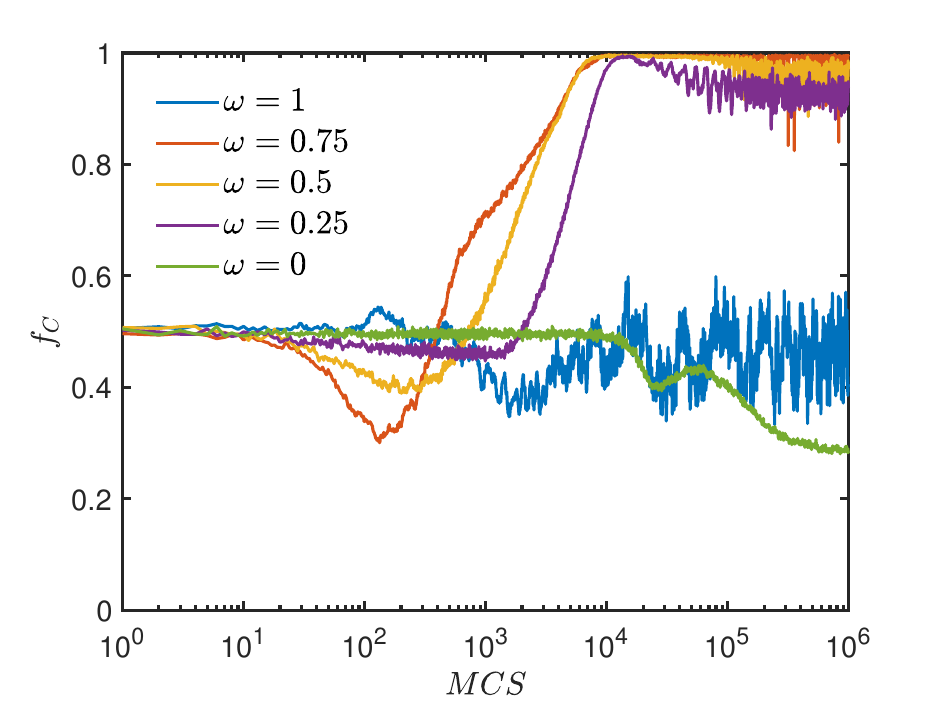}\put(2.5,65){(a)}\end{overpic}
\begin{overpic}[width=0.3\linewidth]{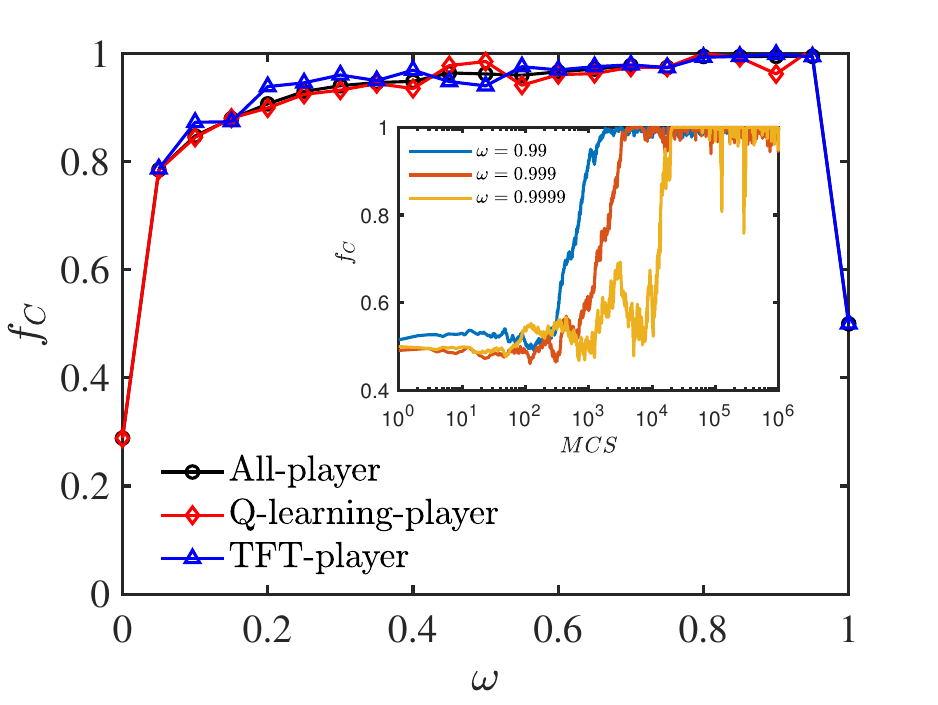}\put(2.5,65){(b)}\end{overpic}
\begin{overpic}[width=0.3\linewidth]{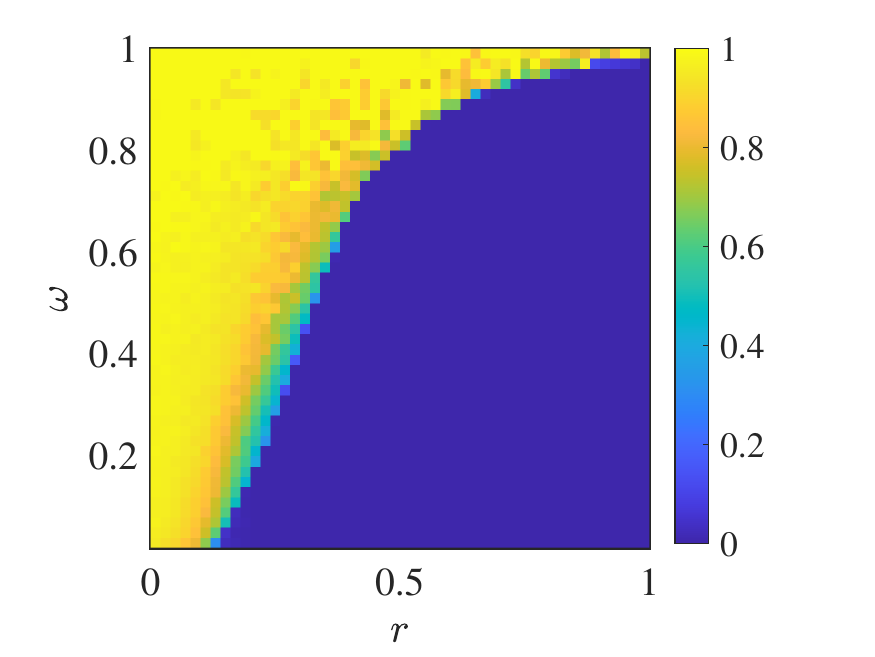}\put(2.5,65){(c)}\end{overpic}
\caption{The evolution of cooperation with PM. $\omega$ is the probability that player evolves according to the TFT rule within each round.
	(a) Time series of cooperation level $f_C$ for several typical $\omega$.
	(b) The prevalence of cooperation $f_C$ versus $\omega$, of all player groups, separately computed for people within TFT mode, Q-learning mode, and for the whole population; the inset provides three time series of $f_C$ for the whole population for three cases of $\omega\rightarrow 1$.
	(c) Heat map for the cooperation prevalence $f_C$ in the $r-\omega $ parameter space.
	The game parameter $r=0.1$ in (a) and (b). 
	Other parameters: $\varepsilon = 0.01$, $\alpha =0.1$, and $\gamma =0.9$.
}
\label{fig:PM}
\end{figure*} 

Note that, the above procedure follows a typical synchronous update process, and each round [steps (2)-(4)] constitutes a Monte Carlo step, where every player updates exactly once. 	
	
The difference in the procedure for PM is within in step (1), where no behavioral mode needs to fixed for the player. Instead, at the beginning of each round $t$, each player randomly selects the TFT mode with the probability $\omega$, otherwise the Q-learning mode is chosen. Notice that, since all players could behave within two modes from time to time, every player will be assigned with a Q-table at the very beginning ($t=0$) in step (1).
	
Compared to PM, the procedure for AM differs in the mode selection. Within step (1), one of the two modes is randomly chosen for initialization, but at the beginning of each new round $t+1$,  player $i$ compares the payoff obtained at present round $\Pi_i(t)$ with the one at the last round $\Pi_i(t-1)$; if $\Pi_i(t)\ge \Pi_i(t-1)$, player $i$ keeps within the old behavioral mode, otherwise player $i$ switches to the other mode. Notice that in both PM and AM, players revise their Q-tables at each time step even though they may not behave in the Q-learning mode to mimic the observed reality where individuals continuously draw lesson from their experiences. The pseudocode for the details of three implementations are given in Appendix A.
	
As can be seen, the probability $\omega$ determines the mixture ratio of the two behavioral modes in SM and PM. As $\omega\rightarrow$ 0 or 1, our mode is recovered to the evolution with pure Q-learning algorithm, or with pure TFT rule~\cite{nowak1992tit,nowak1993strategy}, and we are interested in the mixture case where $0<\omega<1$. By contrast, this mixture ratio is adaptively adjusted in AM implication, where the evolution within either pure mode is rarely seen.
In our practice, we fix $\varepsilon = 0.01$ to allow some degree of random exploration. We compute the cooperation fraction $f_C$ defined as the number of action $C$ divided by $N$ as the primary order parameter to measure the overall preference of the population.

\section{3. Results}\label{sec:results}
\subsection{3.1. Structural mixing}
Let's first examine the impact of mode mixture on the evolution of cooperation with the SM implementation, where $\omega$ is the proportion of players within TFT mode, and they remain in the same mode throughout the evolution. Fig.~\ref{fig:SM} reports the results on the 2d square lattice with SM. As the benchmarks, let's check the two extreme cases (i.e. $\omega = 0, 1$), which corresponds to the single-mode scenarios of Q-learning and TFT mode, respectively. Specifically, when the population is within pure Q-learning mode, the resulting $f_C\approx 0.29$, meaning that players learn to cooperate to some degree but the prevalence is not high. When the population is within pure TFT mode, the resulting cooperation prevalence depends on the initial condition, here in our study $f_C\approx f_C(t=0)\approx 0.5$.

When the two modes are mixed ($0<\omega < 1$), their cooperation prevalence is promoted in the given cases (i.e. $\omega = 0.25, 0.5, 0.75$) compared to the two single-mode scenarios, see Fig.~\ref{fig:SM}(a). In all three mixing cases, $f_C>0.5$ and the prevalence is highest for $\omega=0.75$. Given the pure TFT scenario, this indicates there is a non-monotonic dependence of cooperation prevalence on $\omega$. This is confirmed by plotting the cooperation prevalence versus the mode proportion $\omega$, shown in Fig.~\ref{fig:SM}(b). As seen, the overall prevalence $f_C$ increases with escalating $\omega$, almost full cooperation is reached when $0.8<\omega<1$. Detailed statistical analysis shows the cooperation prevalence is slightly higher for those within the TFT mode than those Q-learners. The trend of $\omega\rightarrow 1$ indicates that the addition of very few Q-learners into the TFT population is sufficient to drive the whole system into full-cooperation state, and this is indeed confirmed by our simulations [the inset within Fig.~\ref{fig:SM}(b)]. This means that, individuals in Q-learning mode play the role of cooperation catalysts, very few of them are able to promote the overall cooperation. 

Fig.~\ref{fig:SM}(c) provides the heat map for the cooperation prevalence in $r-\omega$ parameter domain to further investigate the impact of the game parameter $r$. As expected, $f_C$ is reduced as the dilemma strength $r$ is increased, and the transitions from high cooperation to low cooperation state are all of continuous type.

\begin{figure*}[htbp]
\centering
\begin{overpic}[width=0.4\linewidth]{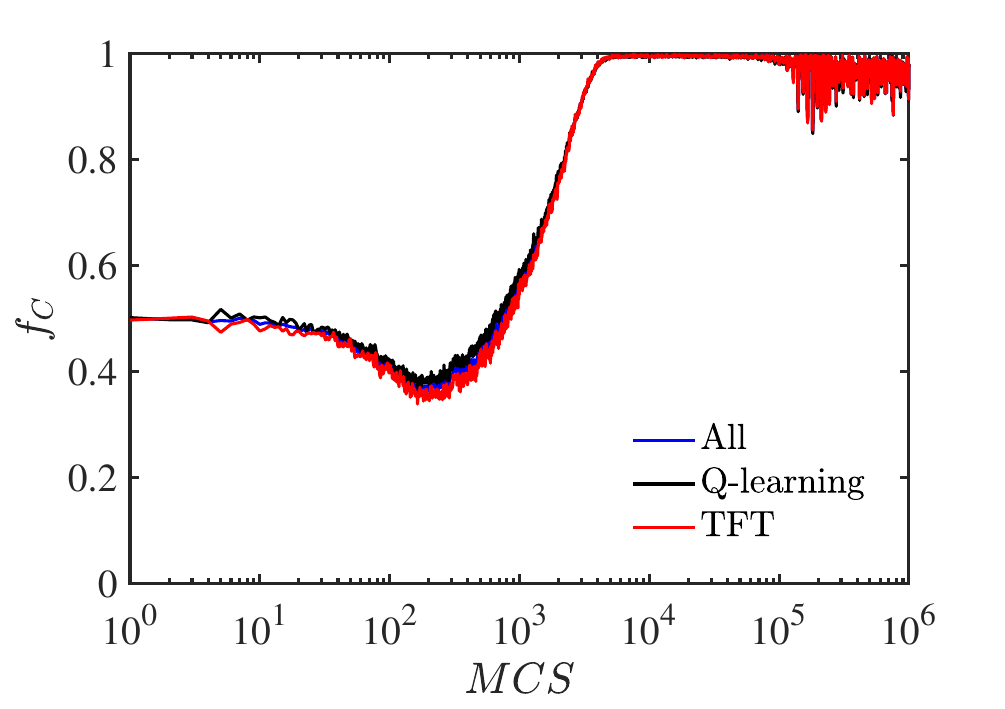}\put(2.5,65){(a)}\end{overpic}
\begin{overpic}[width=0.4\linewidth]{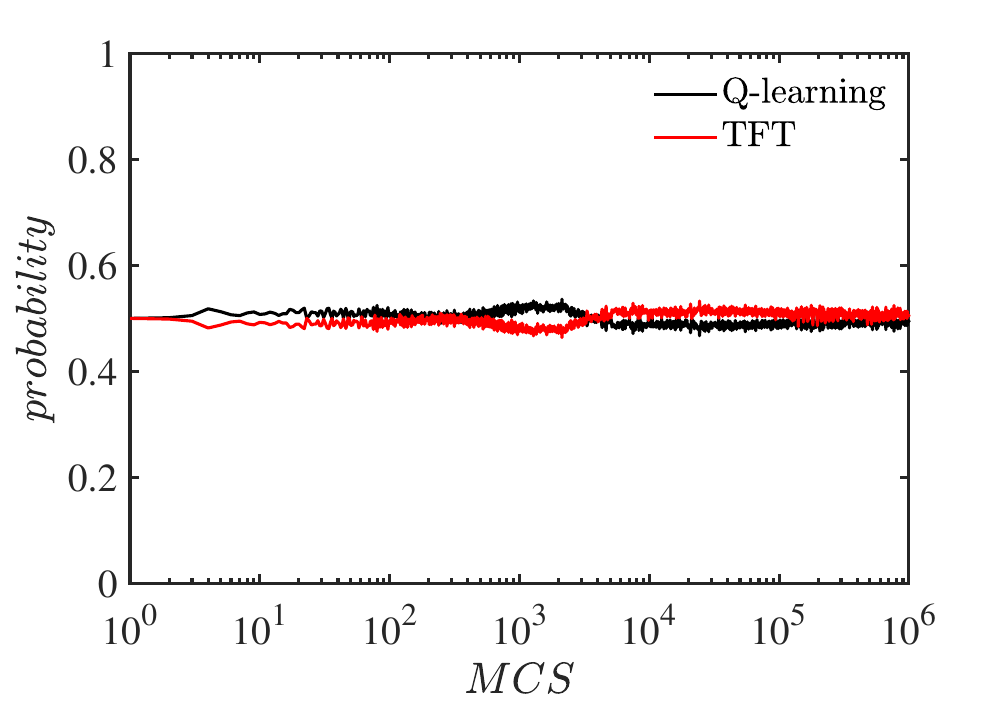}\put(2.5,65){(b)}\end{overpic}
\caption{Time series of cooperation prevalence with AM.
	(a) The prevalence of cooperation $f_C$ computed separately for players with TFT and Q-learning, together with the whole population for comparison.
	(b) The time evolution of probabilities of adopting the two modes by the population.
	Other parameters: $\varepsilon = 0.01$, $\alpha =0.1$, $\gamma =0.9$, and $r=0.1$.
}
\label{fig:AM}
\end{figure*}

\subsection{3.2. Probabilistic mixing}		
Fig.~\ref{fig:PM} shows the impact of the mode mixture on the evolution cooperation with the PM implementation. Bear in mind, here the behavioral mode for each players is switching from time to time, and $\omega$ is interpreted as the probability to behave within the mode of TFT rule at each round.

Compared to Fig.~\ref{fig:SM}, the two single-mode scenarios ($\omega=0, 1$), shows exactly the same prevalence in cooperation, as expected. The difference, however, lies in the observations in the mixture cases, where the cooperation prevalences in all three given cases ($\omega=0.25, 0.5, 0.75$) jump to be very high $f_C\rightarrow 1$ in the long term.
Fig.~\ref{fig:PM}(b) clearly shows this difference compared to Fig.~\ref{fig:SM}(b). We observe that there is a wide of the parameter $\omega$, where the cooperation prevalence $f_C>0.8$. Also, the prevalences computed separately for people within TFT and Q-learning mode show that the two are now statistically the same, no quantitative difference is detected. Once again, when $\omega\rightarrow 1$, the presence of very few Q-learning mode acts as the cooperation catalysts, driving the population into almost full cooperation [see the inset within Fig.~\ref{fig:PM}(b)]. Though, the time series of $f_C$ indicates that the evolution with too few of Q-learning mode requires longer transients to have $f_C\rightarrow 1$.
The heat map shown in Fig.~\ref{fig:PM}(c) shows a widened region for cooperation compared to Fig.~\ref{fig:SM}(c), and the cooperation-defection transition becomes more abrupt.

Note that, the long term evolution (e.g. $t>10^5$) in Fig.~\ref{fig:PM}(a) exhibits some large fluctuations in the cooperation prevalence, though a decent cooperation prevalence is still maintained. These fluctuations are due to the Q-learning players, who turn to defection to obtain higher payoffs thus reduce $f_C$, but the presence of players within TFT mode inhibits this trend. This forces the Q-learning players turn back to cooperation again and the prevalence increase again. This process repeats again and again, and an oscillation is formed. Detailed discussion is provided in Appendix B.

\subsection{3.3. Adaptive mixing}
Different from the above two implementations, where the two mode fractions are predetermined, the mode adoption in AM is adjusted from time to time according to the payoff changes. Fig.~\ref{fig:AM} shows a typical evolution where both the cooperation preference and the probabilities of mode adoption are monitored. Fig.~\ref{fig:AM}(a) shows that after a slight decay in $f_C$ for $t<10^2$, the cooperation prevalence continuously increases till $f_C\rightarrow 1$ is reached around $t>10^4$. Detailed examination shows that the prevalence obtained within Q-learning and TFT are almost identical. Further evolution ($t>10^5$), however, some fluctuations in cooperation appear, with the same observations made in PM [see Fig.~\ref{fig:PM}(a)].

Fig.~\ref{fig:AM}(b) shows that the adoption probability for the two modes remains largely identical. Though a detailed examination shows that the probability of adopting Q-learning mode is slightly larger than TFT mode, and later there is a crossover at around $t=10^3$.

\section{4. Mechanism analysis}\label{sec:results}

\begin{figure*}[htbp]
\centering
\begin{overpic}[width=0.3\linewidth]{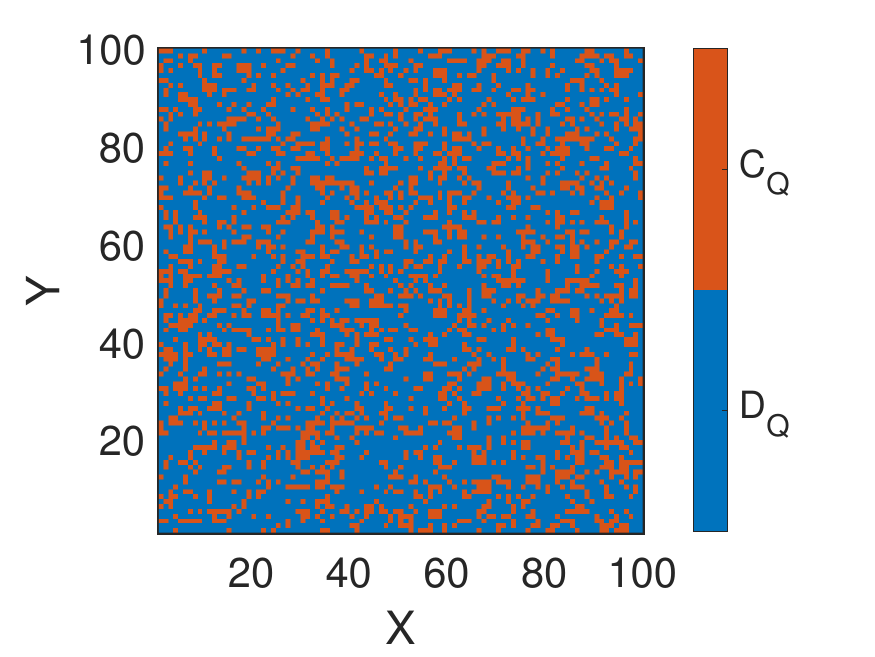}\put(2.5,65){(a)}\end{overpic}
\begin{overpic}[width=0.3\linewidth]{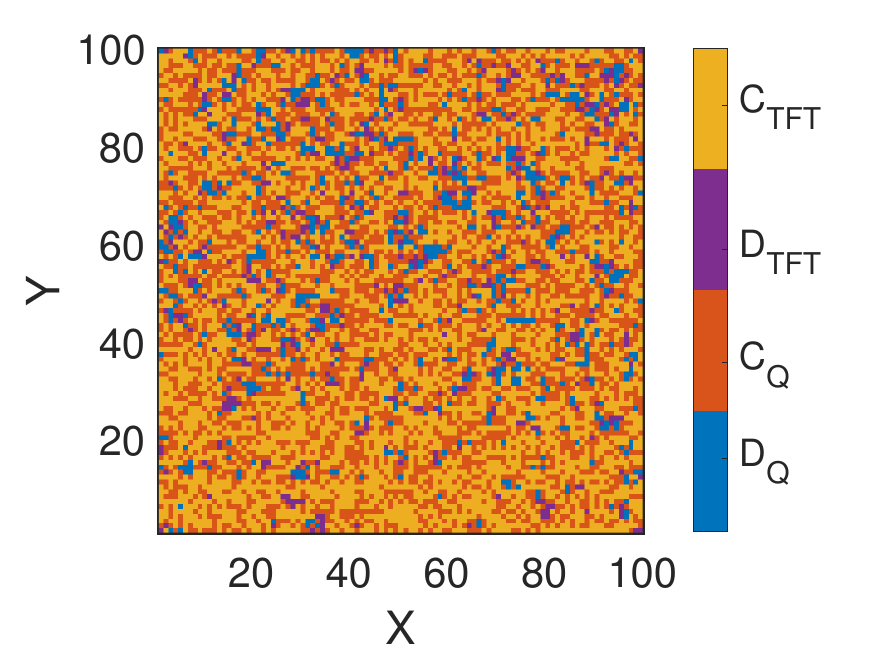}\put(2.5,65){(b)}\end{overpic}
\begin{overpic}[width=0.3\linewidth]{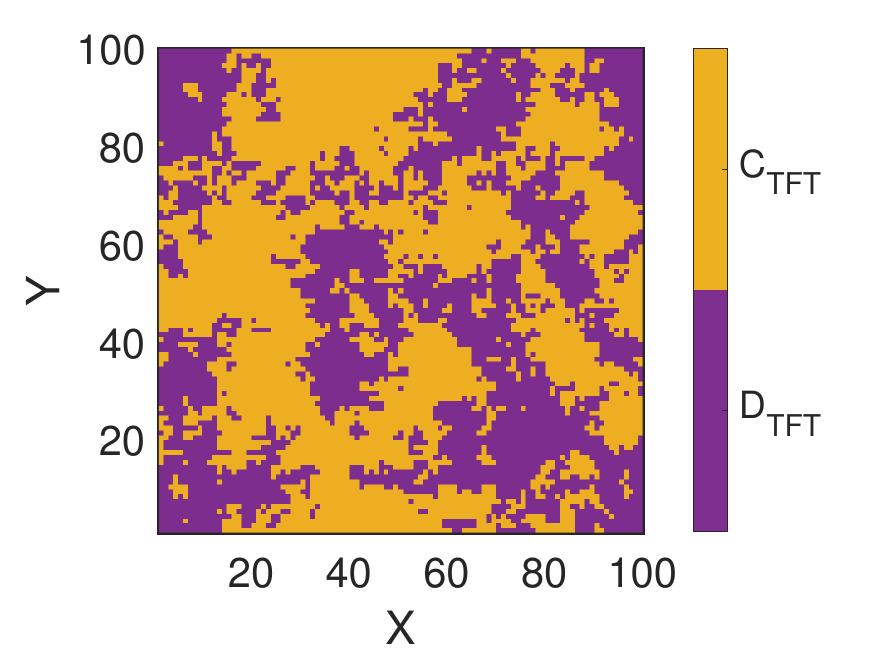}\put(2.5,65){(c)}\end{overpic}
\caption{Typical spatial patterns on $2d$ square lattice with SM at $t=2\times 10^{6} $ for three $\omega$: (a) $\omega=0$, (b) $\omega=0.5$, (c) $\omega=1$.
Both the strategy and mode information are color coded, e.g. blue pixels ($D_{Q}$) denote the players are within the Q-learning mode and adopt defection as their strategies.
$f_C\approx0.29$, 0.86, and 0.51 in (a-c), respectively.
Other parameters: $\varepsilon = 0.01$, $\alpha =0.1$, $\gamma =0.9$, and $r=0.1$.
}
\label{fig:pattern}
\end{figure*}

\emph{Structure mixing (SM)} --- To reveal the mechanism why the mode mixing promotes the cooperation prevalence, let's move back the evolution within the SM implementation. To gain some intuition, let's first examine the spatial pattern of states with different probabilities $\omega$, shown in Fig.~\ref{fig:pattern}. Fig.~\ref{fig:pattern}(a) illustrates that when all players behave within the Q-learning mode  (i.e., $\omega=0 $ ), the majority of players in the group choose strategy $D$, and no obvious clusters are seen for those cooperators. Instead, when the population behave within only TFT mode, the evolution experiences a coarse-graining-like process that forms several large clusters [see Fig.~\ref{fig:pattern}(c)]. The configurations of these clusters are time-varying, but the cooperation prevalence $f_C$ remains at one half on average. The typical mixture case is shown in Fig.~\ref{fig:pattern}(b), where half of the population adopts the TFT mode while the other half behaves with Q-learning ($\omega=0.5 $). The resulting cooperation prevalence $f_C\approx 0.86$, much higher than the value in either single-mode scenario. Different from Fig.~\ref{fig:pattern}(c), no obvious cluster configuration is observed, the pattern is rather similar to Fig.~\ref{fig:pattern}(a). This means that the emergence of cooperation could not be attributed to the network reciprocity, a well-known mechanism for structured population in the imitation learning framework~\cite{Szabo2007evolutionary, Perc2017statistical}. 
	
\begin{figure*}[htbp]
\centering
\begin{overpic}[width=0.4\linewidth]{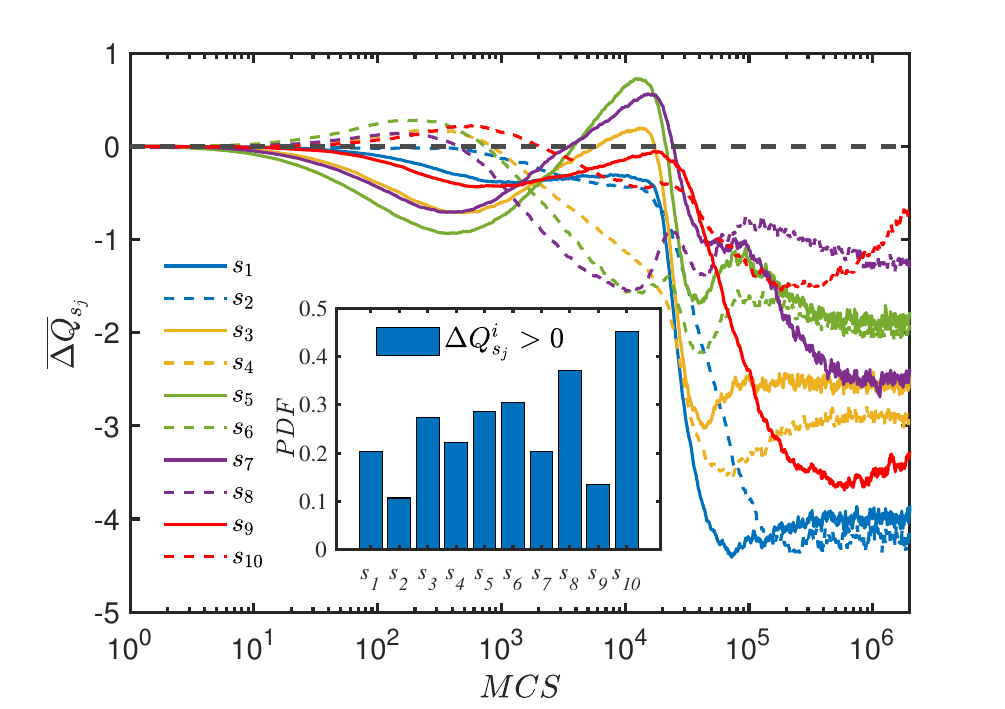}\put(2.5,65){(a)}\end{overpic}
\begin{overpic}[width=0.4\linewidth]{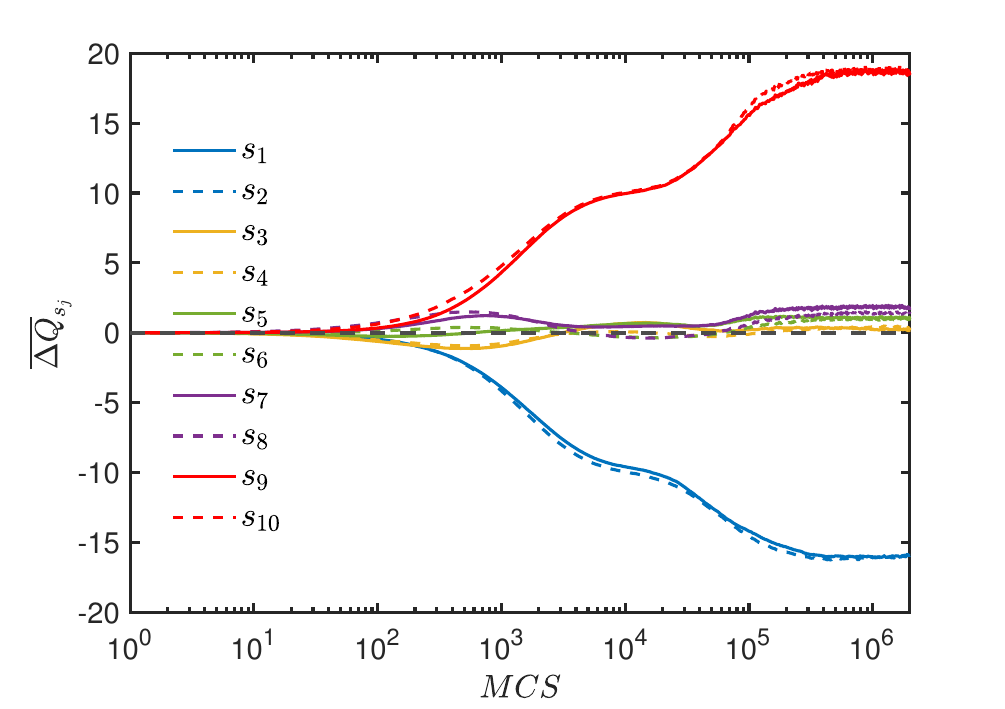}\put(2.5,65){(b)}\end{overpic}
\caption{The temporal evolution of the average 	Q-value difference $\overline{\Delta Q}_{s_j}$ within all 10 states in SM for (a) $\omega=0$ and (b) $\omega=0.5$.
The inset in (a) is the probability density function (PDF) of the Q-learning players with $\Delta Q^i_{s_j} > 0$ in different states at $t=2\times 10^{6}$, where $\Delta Q^i_{s_j}=Q_{s_{j}, C}^{i}-Q_{s_{j}, D}^{i}$. The black dashed line in both plots corresponds to $\overline{\Delta Q}_{s_j}=0$ for reference.
Other parameters: $\varepsilon = 0.01$, $\alpha =0.1$, $\gamma =0.9$, $r=0.1$.
}
\label{fig:dQ_SM}
\end{figure*}
		
To grasp the mechanism, let's turn to the evolution of the Q-tables, which captures the preference evolution of the players within the Q-learning mode. 
Specifically, we define the average Q-value difference within a given state $s_j\in S$ as
\begin{equation}
\begin{aligned}
	\overline{\Delta Q}_{s_{j}}=\frac{1}{N} \sum_{i=1}^{N}\left(Q_{s_{j}, C}^{i}-Q_{s_{j}, D}^{i}\right).\label{con:5}
\end{aligned}
\end{equation} 		
When $\overline{\Delta Q}_{s_j}>0$, players are more inclined to choose cooperation within state $s_j$; conversely, they prefer to defect when $\overline{\Delta Q}_{s_j}<0$. 
As the benchmark, let's first look at the scenario of pure Q-learning mode ($\omega=0$), see Fig.~\ref{fig:dQ_SM}(a). It shows that the $\overline{\Delta Q}_{s_j}$ of all states go up and down, but eventually, all of them become negative $\overline{\Delta Q}_{s_j}<0$ ($i=1,2,...,10$). This means that on average players prefer to defect in all possible states, though the probability density function (PDF) shown [inset of Fig.~\ref{fig:dQ_SM}(a)] indicates that there is still some probability for $\Delta Q^i_{s_j}>0$ at the individual level, which explains the low but non-vanishing cooperation preference in the pure mode of Q-learning. Notice that, the preference for those cooperators (within $s_{1,3,5,7,9}$) and defectors (within $s_{2,4,6,8,10}$) have opposite trends by evolution initially, though all become negative, in particularly for those who are surrounded by defectors are determined to defect where $\overline{\Delta Q}_{s_{1,2}}\rightarrow - 4$. 

The cooperation preference in the typical mixing case is dramatically different, as shown in Fig.~\ref{fig:dQ_SM}(b). As can be seen, the average Q-value differences $\overline{\Delta Q}_{s_j}$ initially are close to zero, meaning that no obvious preference is developed when they are unfamiliar with the surroundings. After the transient ($t \gtrsim 10^4$), three kinds of preferences are formed as clearly shown in $\overline{\Delta Q}_{s_j}$. For those surrounded by four cooperators ($s_{9,10}$), they reciprocate with C as $\overline{\Delta Q}_{s_{9,10}}>15$ in the end; instead for those surrounded by four defectors ($s_{1,2}$), they reciprocate with D as $\overline{\Delta Q}_{s_{1,2}}<-15$. For the rest cases, $\overline{\Delta Q}_{s_{3,...,8}}>0$ means that they all prefer cooperation on average, and detailed examination shows that the more cooperators in the neighborhood, the stronger preference in cooperation (with a larger value of $\overline{\Delta Q}_{s_j}$).
	
\begin{figure*}[htbp]
\centering
\begin{overpic}[width=0.3\linewidth]{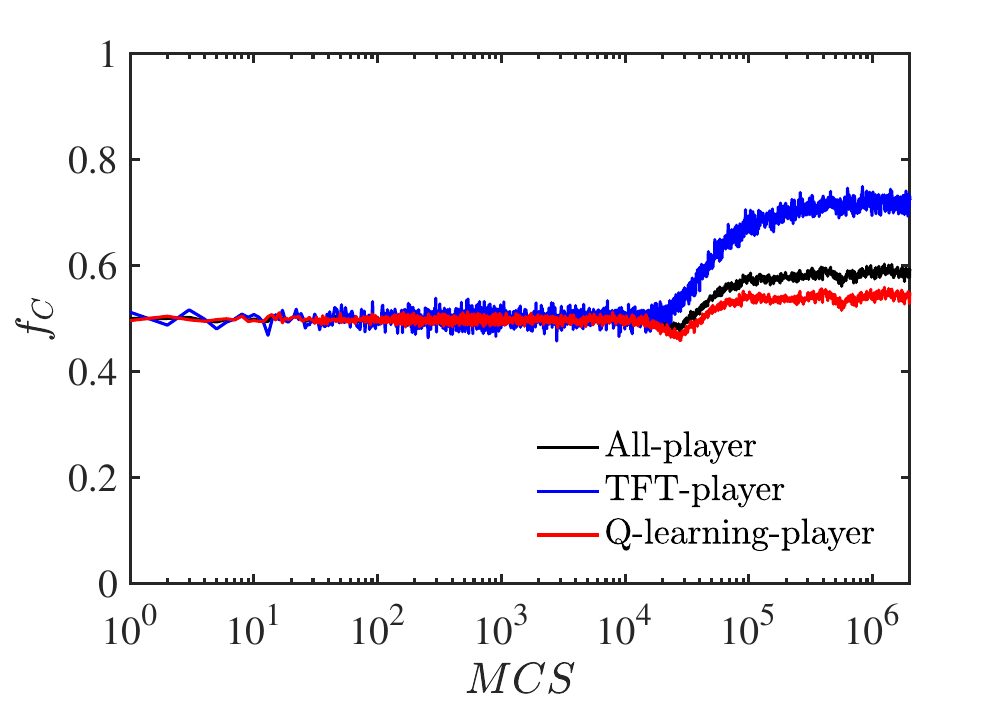}\put(2.5,65){(a)}\end{overpic}
\begin{overpic}[width=0.3\linewidth]{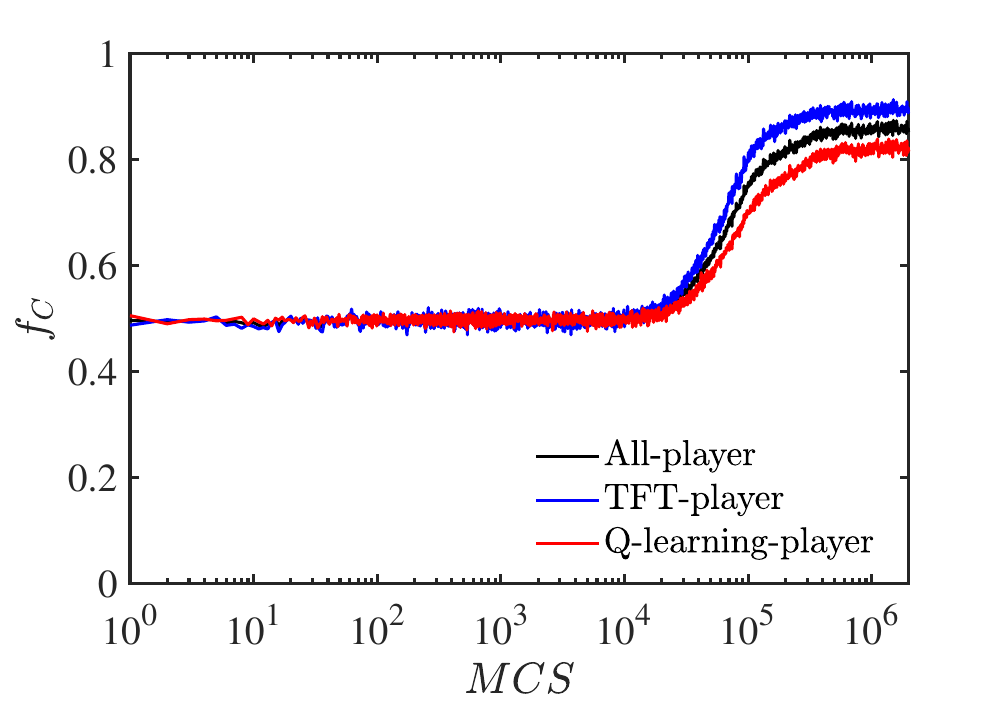}\put(2.5,65){(b)}\end{overpic}
\begin{overpic}[width=0.3\linewidth]{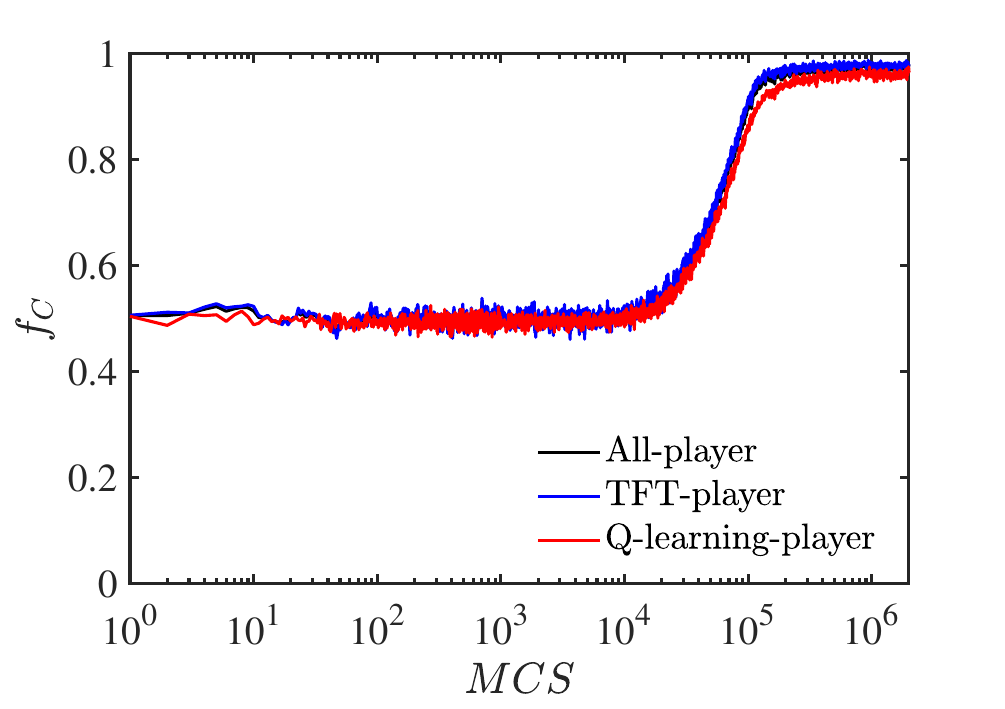}\put(2.5,65){(c)}\end{overpic}\\
\begin{overpic}[width=0.3\linewidth]{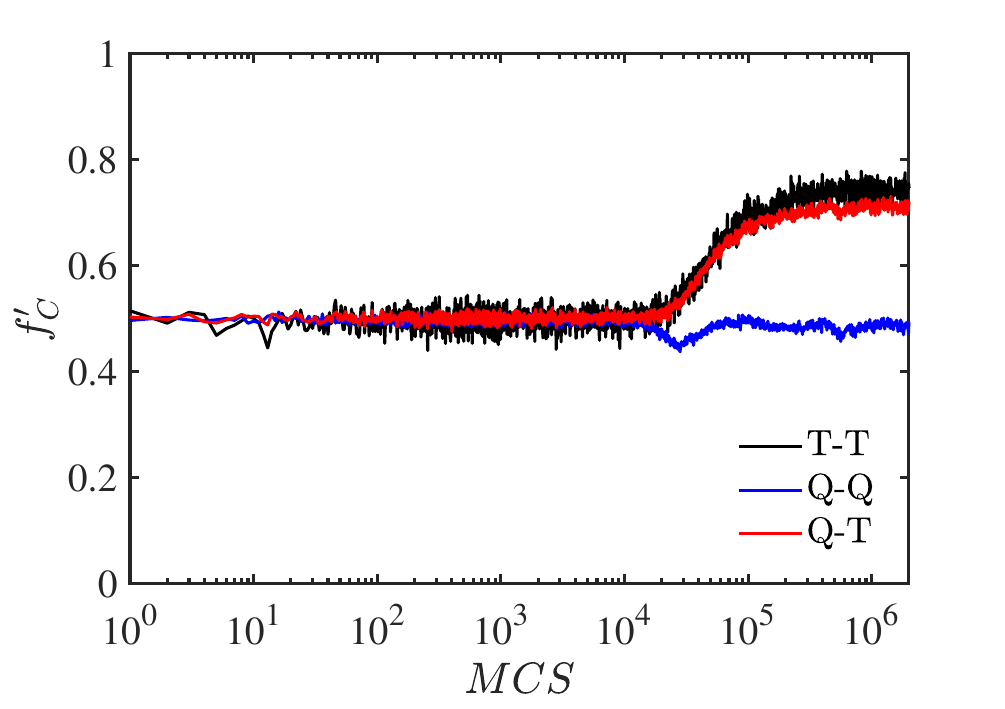}\put(2.5,65){(d)}\end{overpic}
\begin{overpic}[width=0.3\linewidth]{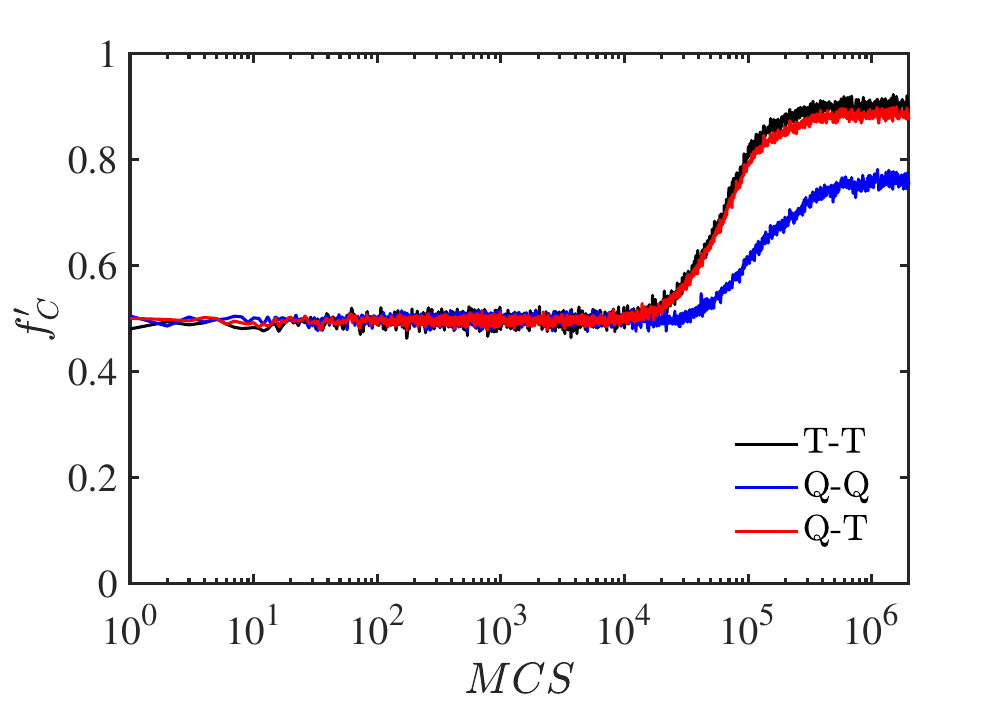}\put(2.5,65){(e)}\end{overpic}
\begin{overpic}[width=0.3\linewidth]{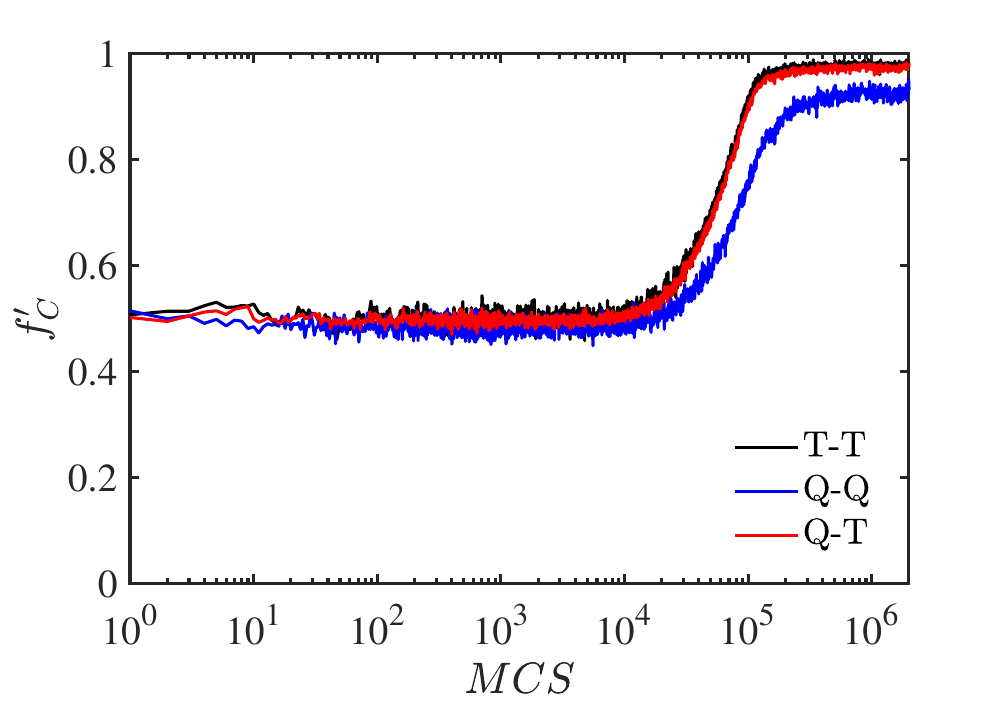}\put(2.5,65){(f)}\end{overpic}
\caption{Time series of the two kinds of cooperation prevalences with SM for different $\omega$: (a, d) $\omega=0.25$, (b, e) $\omega=0.5$, (c, f) $\omega=0.75$.
	The first row (a-c) are the cooperation prevalence computed separately for players within the TFT, Q-learning mode, and the whole population.
	The second row (d-f) are the cooperation propensity of the players who are within TFT versus TFT, TFT versus Q-learning, Q-learning versus Q-learning.
	Other parameters: $\varepsilon = 0.01$, $\alpha =0.1$, $\gamma =0.9$, $r=0.1$.
}
\label{fig:ts_SM}
\end{figure*}

To further unveil why mixing promotes cooperation, we show the time series of the cooperation prevalence $f_C$ of players in TFT and Q-learning mode separately for three typical probabilities ($\omega=0.25, 0.5,0.75$), and compare them to the overall prevalence, shown in Fig.~\ref{fig:ts_SM}(a-c). The results indicate that these prevalences are actually different. In particular, when the players within the Q-learning mode dominate (e.g. $\omega=0.25$), the value of $f_C$ is obviously larger than the one of Q-learning, though the difference is vanishing as $\omega$ becomes larger. This means the distinct role of two behavioral modes in the evolution.

\begin{figure}[tbp]
\centering
\includegraphics[width=1.0\linewidth]{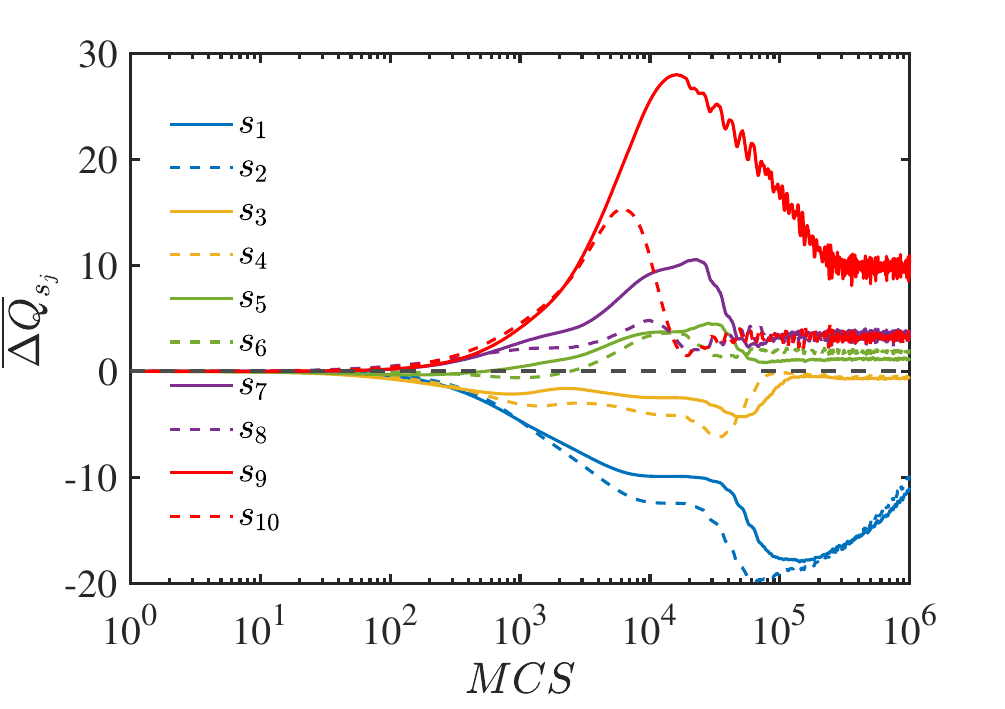}
\caption{The temporal evolution of the average Q-value difference $\overline{\Delta Q}_{s_j}$ within all 10 states in PM for $\omega=0.25$. The black dashed line corresponds to $\overline{\Delta Q}_{s_j}=0$ for reference.
	     Other parameters: $\varepsilon = 0.01$, $\alpha =0.1$, $\gamma =0.9$, $r=0.1$.
}
\label{fig:dQ_PM}
\end{figure}

Since two behavioral modes are present in the population, there are three types of interaction by combination:
 
(i) \emph{Q-learning versus Q-learning} (or Q-Q in short) interactions: when only Q-learners play against Q-learners, they prefer to defect by learning on average, and only a low level of cooperation is achieved ($f_C\approx 0.29$) according to the analysis of Fig.~\ref{fig:dQ_SM}(a).

(ii) \emph{TFT versus TFT} (T-T) interactions: since TFT players only repeat what their neighbors did to them, the population experiences a coarse-graining-like process and forms several clusters with the cooperation prevalence being approximately equal to the initial value [see Fig.~\ref{fig:pattern}(c)]. In our case, $f_C\approx 0.5$ if there are only TFT players.

(iii) \emph{Q-learning versus TFT } (Q-T) interactions: when Q-learners encounter TFT players, they are able to learn that C is a better choice to maximize the rewards, since the action choice of D immediately leads to revenge from their TFT neighbors. This is almost guaranteed if there are two or more TFT neighbors. The confirmation is given by a detailed analysis of $\overline{\Delta Q}_{s_j}$ for Q-learners surrounded by different TFT players in Appendix C.  

Put together, the mechanism behind the promotion of cooperation in mode-mixing population is as follows: initially, for the domains where Q-Q or T-T interactions are dense the cooperation prevalence $f_C$ keeps at the value $f_C(t=0)\approx 0.5$; the promotion comes from the Q-T interactions, where Q-learners learn to be cooperators, and as a result the TFT counterparts then also reciprocate by being cooperators. The high cooperation prevalence within Q-T interactions then triggers a series of propagation of C through T-T interactions (ii), and the same level of $f_C$ is expected for type (ii) and (iii) is expected. This is confirmed in  Fig.~\ref{fig:ts_SM}(d-f). However, the prevalence in Q-Q pair is qualitatively lower, and this is because the cooperation prevalence in type (i) is lower and their cooperation prevalence is to difficult to improve.  

The nature difference in type (i-iii) then also explains why the cooperation prevalence for players in TFT mode is higher than those with Q-learning. This is because the promotion originated from T-Q interactions fail to propagate to the Q-Q interactions, resulting in a relatively lower $f_C$ in Q-learning players. The overall picture also explains the dependence of $f_C$ on the parameter $\omega$ seen in Fig.~\ref{fig:SM}: the larger proportion of TFT players $\omega$ leads to less Q-Q interactions, whereby a higher cooperation prevalence $f_C$ is expected. Notice that, since Q-T interaction is the source of cooperation promotion, a small amount of players need to play within the Q-learning mode, playing the role of catalyst to trigger the cooperation promotion for the whole population.
		
			
\emph{Probabilistic mixing (PM)} --- The above mechanism for SM is still working for explaining the cooperation promotion in the PM implementation, where Q-T interactions still account for the source of the promotion. The difference in PM is that, Q-Q interactions are now hard to persist for any given paired players, as the Q-Q interactions are destroyed from time to time due to the probabilistic mode switching in PM. As a consequence, there is less chance for Q-Q interactions to persist, and the cooperation prevalence is thus driven by either Q-T or propagates along T-T interactions, finally a high level of $f_C$ is seen [see Fig.~\ref{fig:PM}]. And since $\omega$ only controls the probability for the two modes, the specific value for $0<\omega<1$ does not influence the resulting $f_C$ as no Q-Q interactions in SM are present for the given $r$. Though,  Fig.~\ref{fig:PM}(c) indicates that for the case of larger $r$, a higher $\omega$ is required for a decent level of cooperation, as this increases the interaction of type (ii) but reduces the interactions of type (i), and keeps some interactions of type (iii). 

In fact, we can also monitor the evolution of cooperation preference by computing the average Q-value difference $\overline{\Delta Q}_{s_j}$, as shown in Fig.~\ref{fig:dQ_PM}. It shows that the preference in six states $\overline{\Delta Q}_{s_j}>0$ ($j=5,6,7,8,9,10$), while the rest four $\overline{\Delta Q}_{s_j}<0$ ($j=1,2,3,4$), this division becomes most prominent at round $10^4$. Compared to Fig.~\ref{fig:dQ_SM}(b), both show the similar reciprocity property: cooperate when all neighbors cooperate, while defect when all neighbors defect. Afterwards, however, there are declines all absolute values of $\overline{\Delta Q}_{s_j}$, this is especially true for $\overline{\Delta Q}_{s_{9,10}}$. This means players within Q-learning mode may attempt to defect to achieve higher rewards, but the presence of TFT mode immediately feedback defection to their action. This inhibits the decline of $\overline{\Delta Q}_{s_j}$, and finally both values of $\overline{\Delta Q}_{s_{9,10}}$ remain positive to guarantee that the choice of C is statistically more preferred.

\begin{figure}[tbp]
\centering
\includegraphics[width=1.0\linewidth]{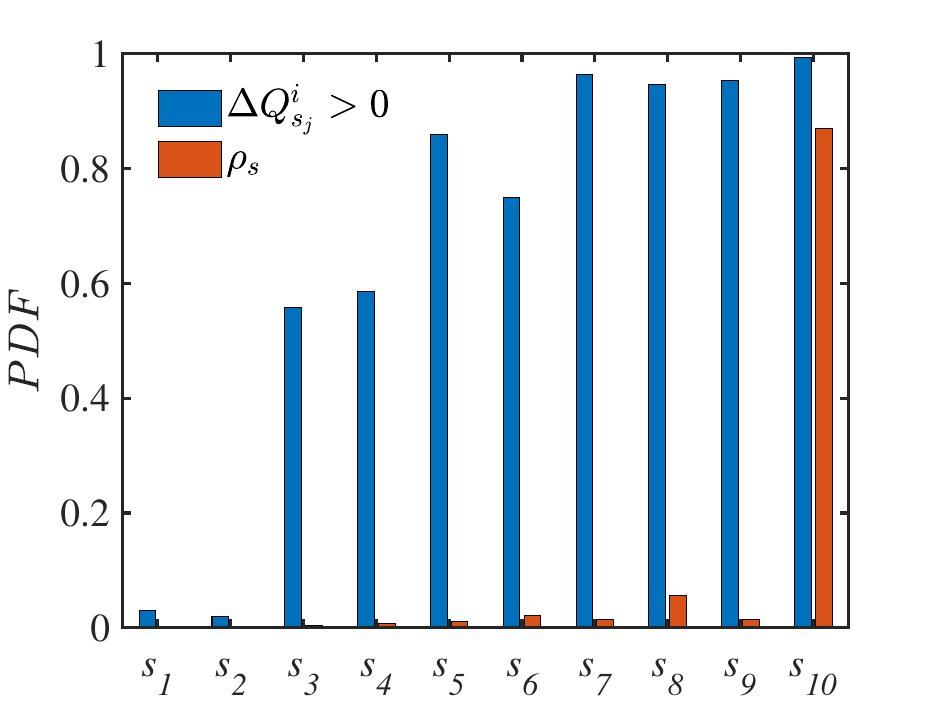}
\caption{The PDF of players' cooperation preference (i.e. the probability for $\Delta Q^i_{s_j}=Q_{s_{j}, C}^{i}-Q_{s_{j}, D}^{i}>0$) within all 10 state, and the density of these states $\rho_s$ per se. Data is sampled from the last $10^4$ before the evolution of $10^{6}$ steps is completed.
Other parameters: $\varepsilon = 0.01$, $\alpha =0.1$, $\gamma =0.9$, $r=0.1$.
}
\label{fig:dQ_PDF_AM}
\end{figure}
	
\emph{Adaptive mixing (AM)} --- Compared to the PM implementation, the logic of mode adoption differs only in the details, i.e. the mode is not adopted probabilistically but  according to a predetermined rule given in model part. From the interaction point of view, the mechanism behind the cooperation prevalence is essentially the same as PM.

To precisely characterize the cooperation prevalence, we sample the value of $\Delta Q^i_{s_j}$ at the individual level defined as $\Delta Q^i_{s_j}=Q_{s_{j}, C}^{i}-Q_{s_{j}, D}^{i}$, where $\Delta Q^i_{s_j}>0$ means that player $i$ prefers $C$ as its action within state $s_j$, otherwise $D$ is favored. 
Fig.~\ref{fig:dQ_PDF_AM} shows the percentage of $\Delta Q^i_{s_j}>0$ in all 10 states. We see that very low probabilities are seen in $s_{1,2}$, meaning players choose to defect when they are surrounded by defectors. By contrast, when the surroundings become cooperative, e.g. $s_{9,10}$, the probability to cooperate becomes high. For most states (e.g. $s_{5-10}$), the cooperation probability is larger than 0.6. In fact, Fig.~\ref{fig:dQ_PDF_AM} also shows that the state $s_{10}$ is dominating, where the density $\rho_{s_{10}}$ is nearly as high as $90\%$, meaning that players are mostly located in a very cooperative surrounding, whereby a decent level of cooperation is expected since the probability of $\Delta Q^i_{s_{10}}> 0$ goes to be 1.

\section{5. Conclusion and discussion}\label{sec:discussion}
	
Motivated by the behavior multimodality in the real world, here we develop a toy model of mixing Q-learning and TFT rules. While Q-learning is a typical RL algorithm, the decision-making is based upon the learned experience, TFT is a well-known reactive strategy. The combination of these two is not contradictory, but complementary. 
In the first implementation (SM), where each player randomly chooses one mode and sticks to it, we reveal that the mode mixture is able to promote and a higher proportion of players in TFT mode generally yields a higher level of cooperation. 
In the second implementation (PM), players behave probabilistically in the either mode in each round, the cooperation promotion is even more pronounced than the case of SM, where nearly full cooperation is expected for a wide range of mixing probability. 
In the third implementation (AM), players are allowed to adopt both modes adaptively, where also nearly full cooperation is seen and the two modes are adopted by the population with approximately equal chance. 

Mechanism analysis shows that in all three implementations, players in Q-learning modes play the role of cooperation catalyzer, where they choose to cooperate with their TFT neighbors and ultimately drive the whole population to a high cooperation level. Since Q-tables reflect the behavioral preference of individuals and exactly captures the thoughts in players' mind, monitoring the statistical properties of Q-tables thus provides important information about the psychologic evolution of the population. This is crucial for understand the logic of human behaviors, and is the great advantage of Q-learning algorithm, which is lacking in the previous framework of social learning~\cite{Szabo2007evolutionary, Perc2017statistical}.
		
Given the nontrivial findings obtained in our work, we hope that efforts could be made to infer the different types of learning from the behavioral experiments. This is an important step towards understanding human behaviors, including cooperation, fairness, honesty, and so on.  
	
\section{Acknowledgments}
This work was supported by the National Natural Science Foundation of China [Grants Nos. 12075144,12165014], and Fundamental Research Funds for the Central Universities (GK202401002). 
		
\appendix

\section{Appendix A: The pseudocode for the three implementations}\label{sec:appendixA}
\begin{figure*}[htbp]
\centering
\begin{overpic}[width=0.3\linewidth]{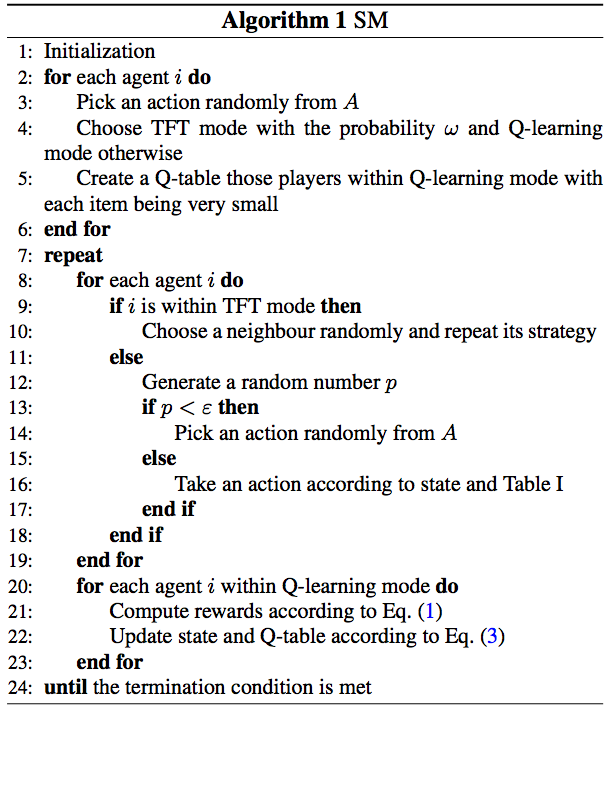} \end{overpic}
\begin{overpic}[width=0.3\linewidth]{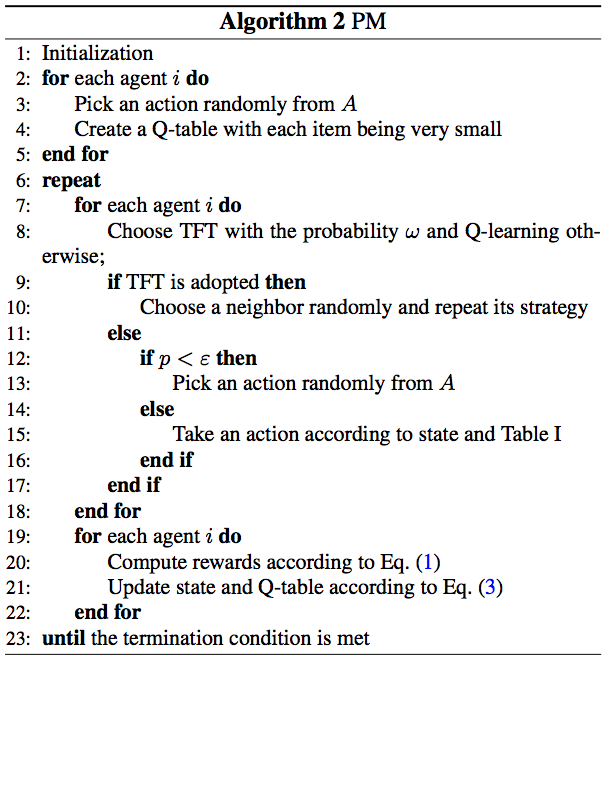} \end{overpic}
\begin{overpic}[width=0.3\linewidth]{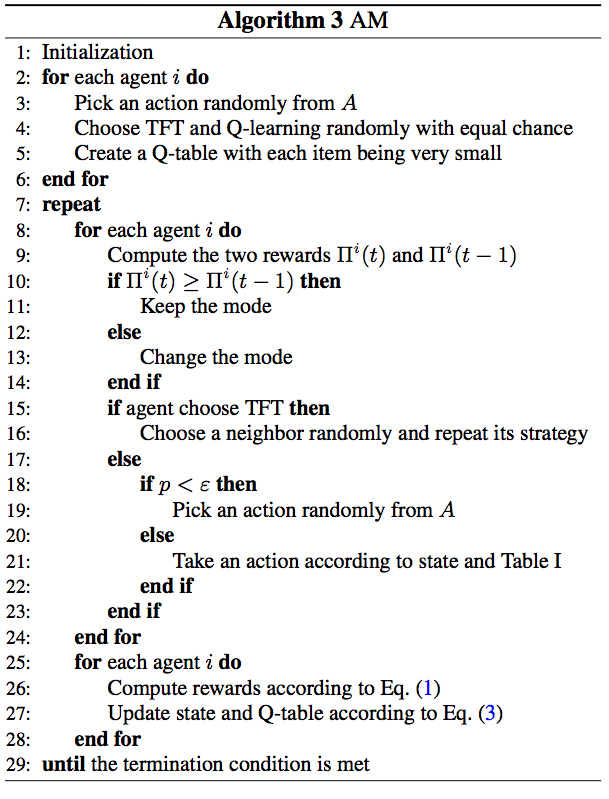} \end{overpic}
\caption{The procedure for our three implementations, where Algorithm 1, 2, 3 are the pseudocodes for SM, PM, and AM, respectively.}
\label{fig:algorithm}
\end{figure*}

To provide detailed procedure of how the three implementations are realized, the pseudocode for SM, PM, and AM is shown below [see Fig.~\ref{fig:algorithm}], respectively.

\section{Appendix B: The phenomena of cooperation oscillation}\label{sec:appendixB}
\begin{figure}[htbp]
\centering
\includegraphics[width=0.8\linewidth]{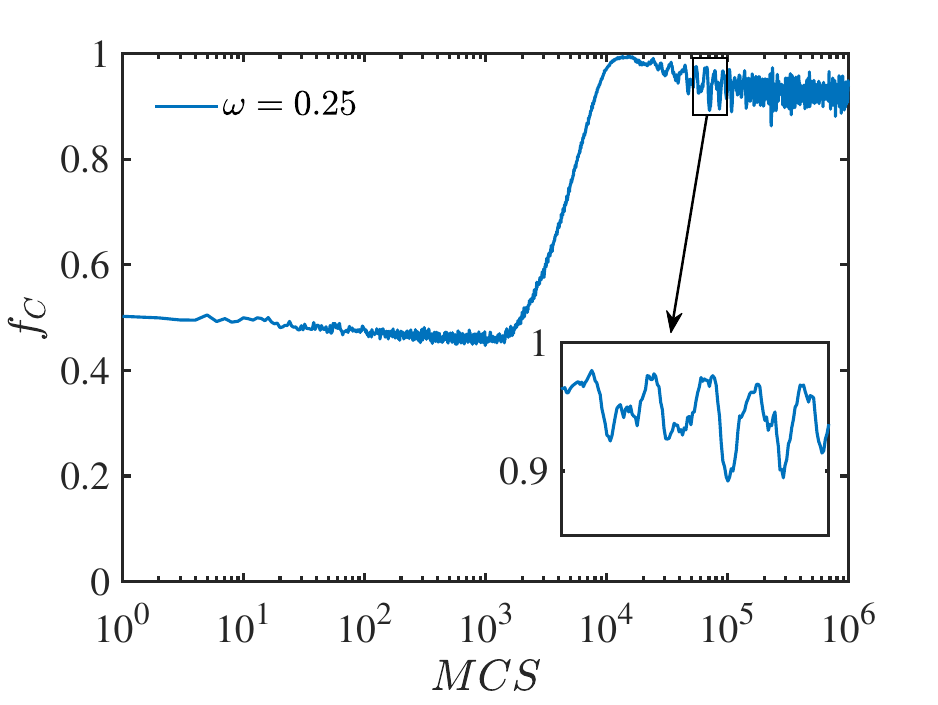}
\caption{ Time series of cooperation level $f_C$ at $\omega=0.25$. The inset is a blowup of the time series in which the oscillatory phenomenon starts.
Other parameters: $\varepsilon = 0.01$, $\alpha =0.1$, $\gamma =0.9$, $r=0.1$.
}
\label{fig:ts_oscillation}
\end{figure}

\begin{figure*}[htbp]
\centering
\begin{overpic}[width=0.3\linewidth]{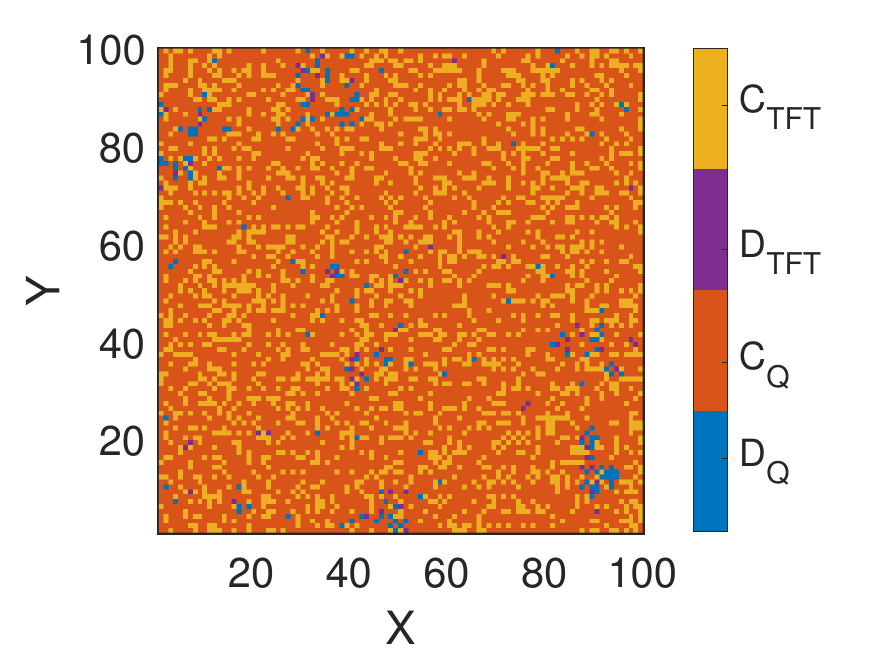}\put(2.5,67){(a)}\end{overpic}
\begin{overpic}[width=0.3\linewidth]{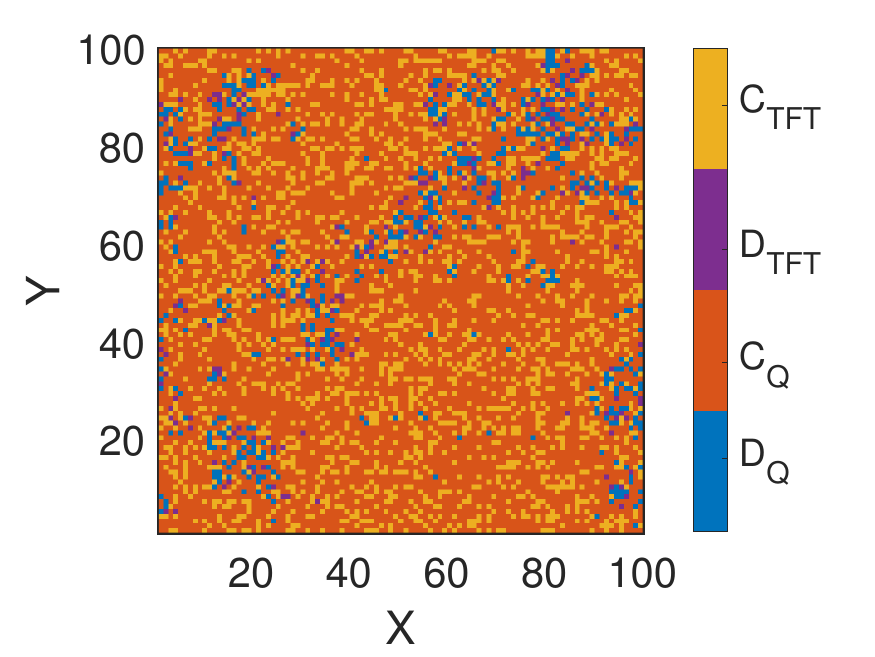}\put(2.5,67){(b)}\end{overpic}
\begin{overpic}[width=0.3\linewidth]{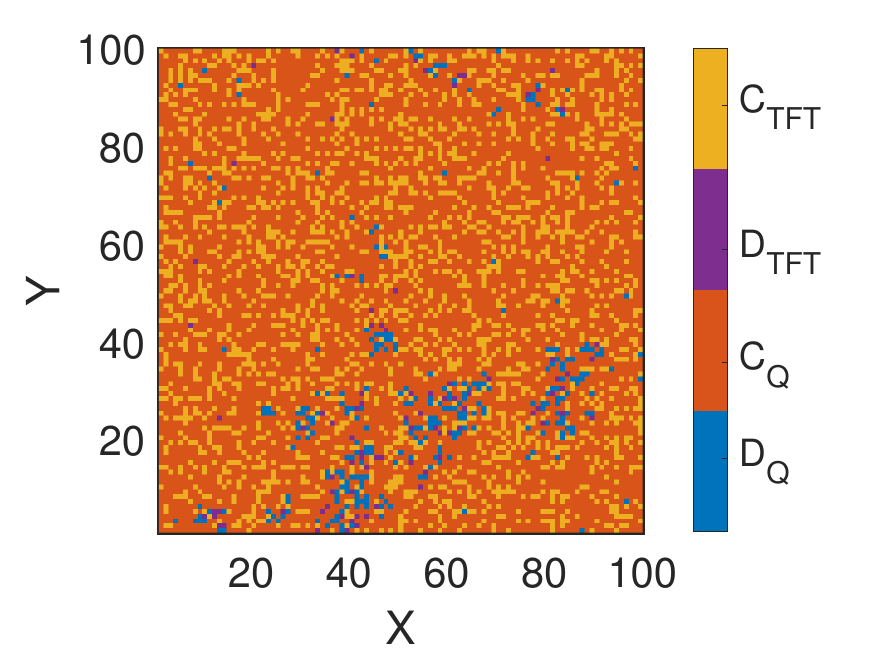}\put(2.5,67){(c)}\end{overpic}
\begin{overpic}[width=0.3\linewidth]{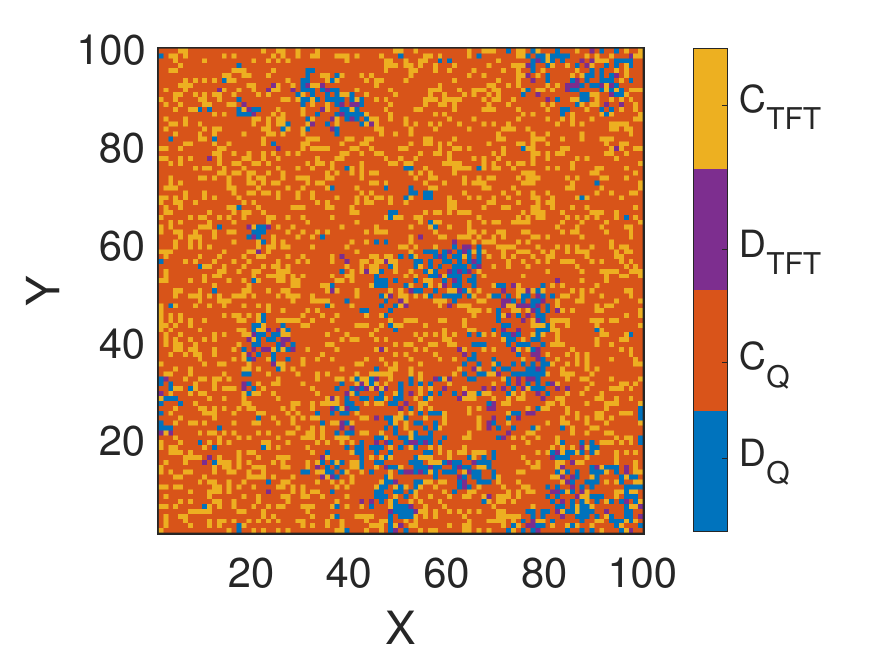}\put(2.5,67){(d)}\end{overpic}
\begin{overpic}[width=0.3\linewidth]{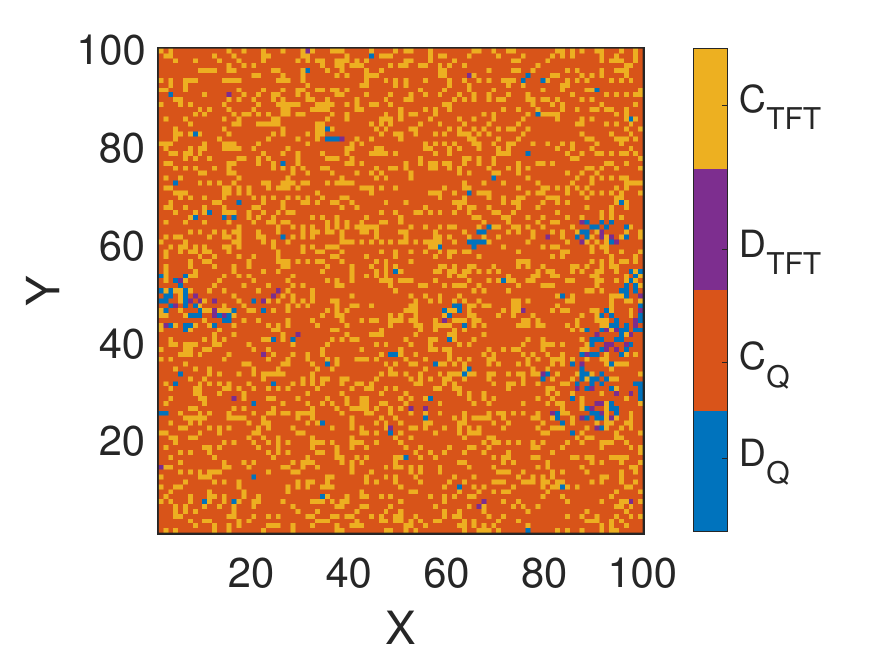}\put(2.5,67){(e)}\end{overpic}
\begin{overpic}[width=0.3\linewidth]{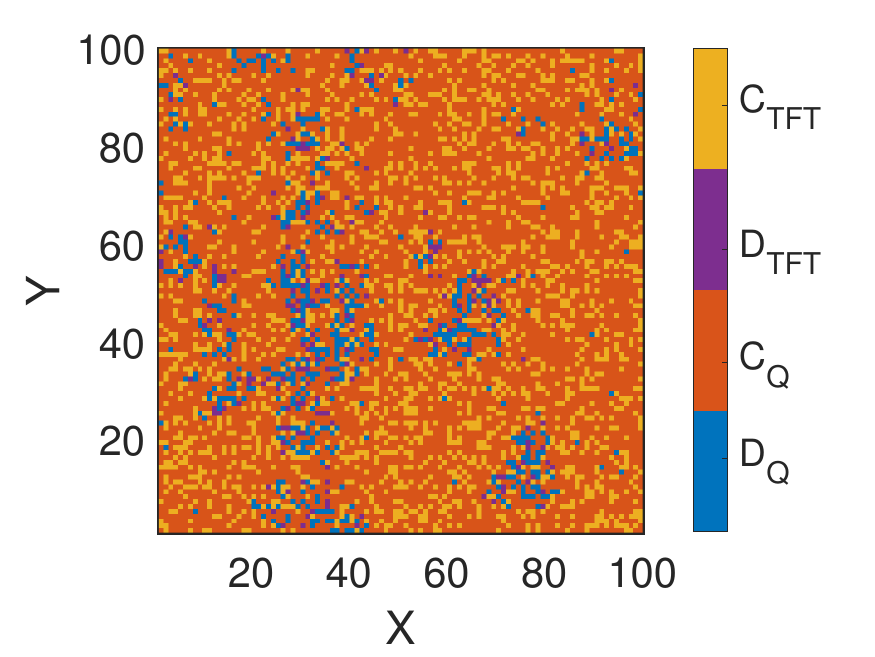}\put(2.5,67){(f)}\end{overpic}
\caption{Typical spatial patterns on $2d$ square lattice with PM for several time steps: (a) $t=43000$, (b) $t=50000$, (c) $t=53000$, (d) $t=67000$, (e) $t=71000$, (f) $t=78000$.
Both the strategy and mode information are color coded, e.g. blue pixels ($D_{Q}$) denote the players are within the Q-learning mode and adopt defection as their strategies.
Other parameters: $\varepsilon = 0.01$, $\alpha =0.1$, $\gamma =0.9$, and $r=0.1$.
}
\label{fig:oscillation_pattern}
\end{figure*}

To examine the large fluctuations present in the long term evolution, Fig.~\ref{fig:ts_oscillation} provides a more clear illustration, where oscillatory fluctuations are seen in the inset. To develop an intuition of how the oscillation is formed, some typical snapshots are shown in Fig.~\ref{fig:oscillation_pattern}. As seen, the fluctuations all come from the players in Q-learning mode, who turn from the choice of C to D. This is triggered by the exploration events which indeed bring them higher payoffs in short term, and thus the number of defective Q-learning players increases [Fig.~\ref{fig:oscillation_pattern}(a) to Fig.~\ref{fig:oscillation_pattern}(b)]. As the surroundings become more defective, other Q-learners tend also to defect, and thus defectors appear in the clustered form. However, as defective Q-learning players increases, their payoffs are continually decreased, and this decreasing trend is strengthened by their TFT neighbors as them immediately also choose D. This lowered reward trend reduces the corresponding Q-values of action C, and therefore those defective players in Q-learning mode turn back to C to maximize their payoffs. But as the surrounding goes back to a high cooperation level [e.g. Fig.~\ref{fig:oscillation_pattern}(c)], the exploration events again cause the decline of $f_C$ and the process repeats again and again.

\section{Appendix C: The interaction of Q-learning versus TFT}\label{sec:appendixC}
\begin{figure*}[htbp]
\centering
\begin{overpic}[width=0.3\linewidth]{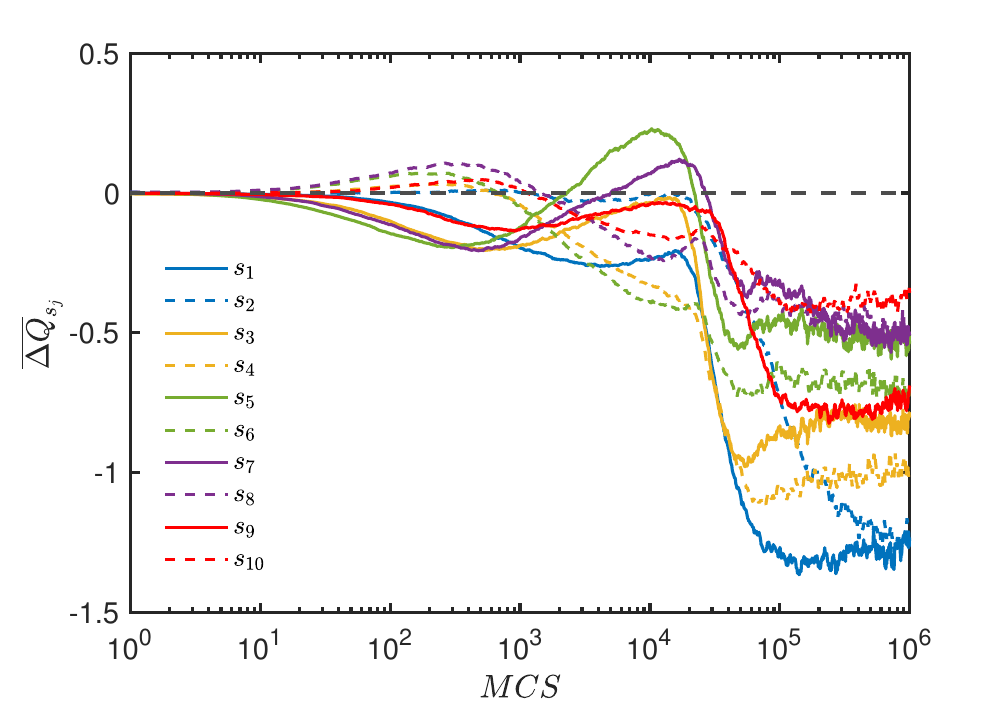}\put(2.5,67){(a)}\end{overpic}
\begin{overpic}[width=0.3\linewidth]{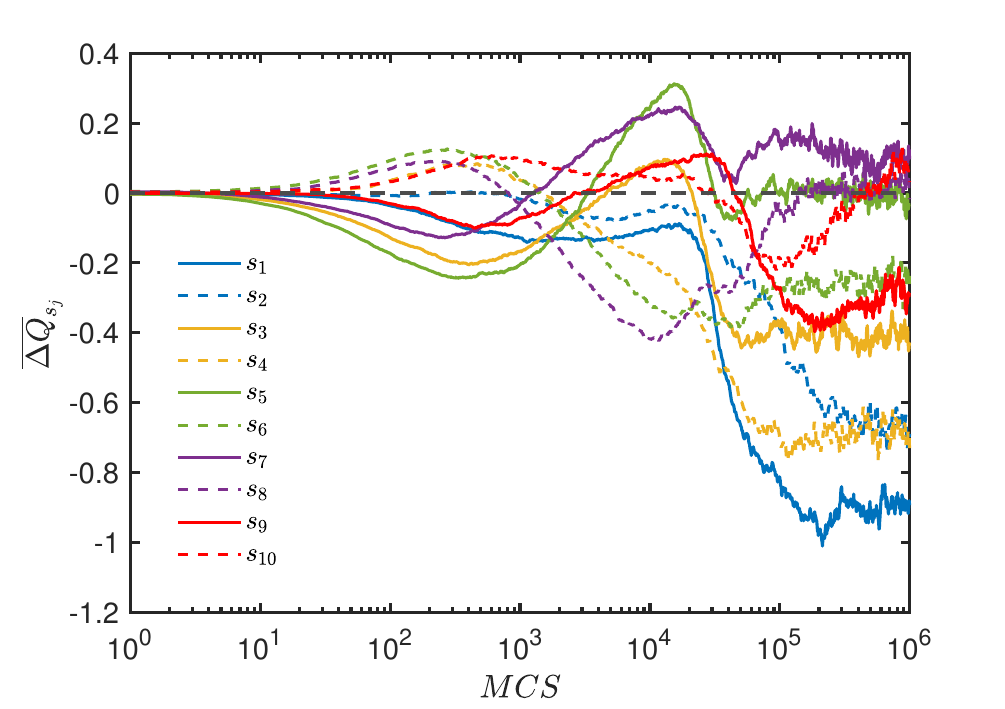}\put(2.5,67){(b)}\end{overpic}
\begin{overpic}[width=0.3\linewidth]{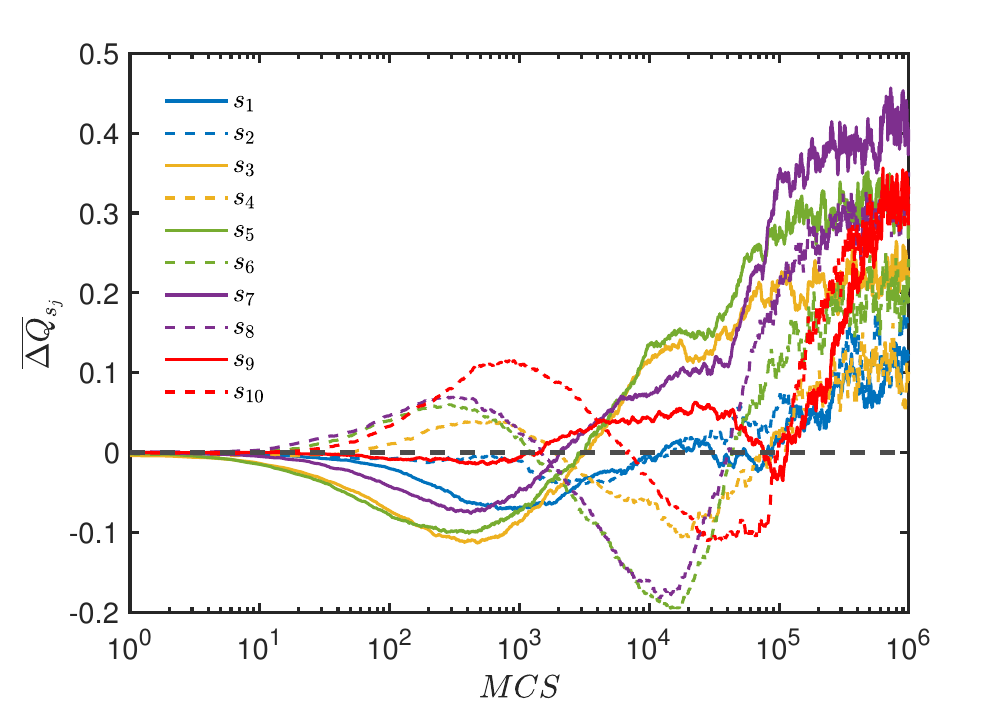}\put(2.5,67){(c)}\end{overpic}
\begin{overpic}[width=0.3\linewidth]{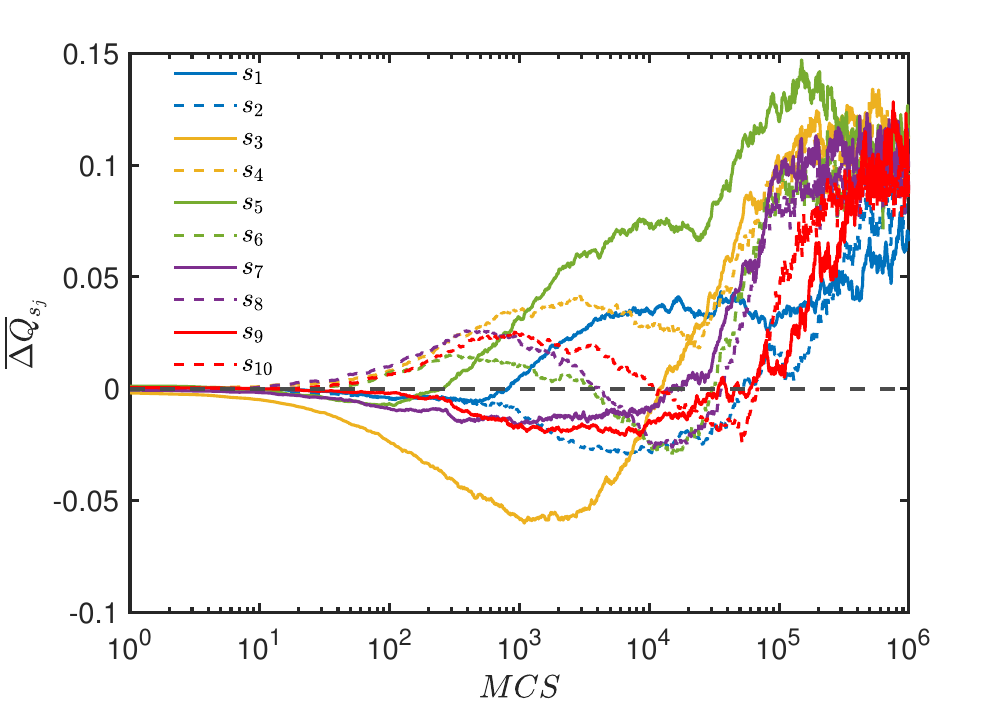}\put(2.5,67){(d)}\end{overpic}
\begin{overpic}[width=0.3\linewidth]{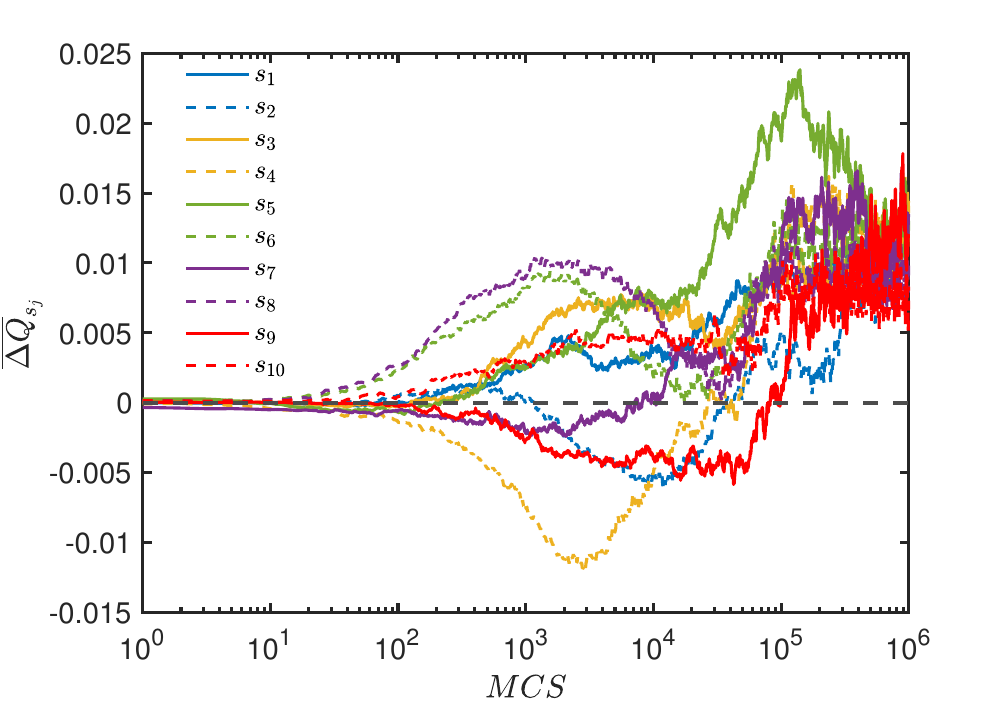}\put(2.5,67){(e)}\end{overpic}
\caption{The temporal evolution of the average 	Q-value difference $\overline{\Delta Q}_j$ within all 10 states in SM for $\omega=0.25$. 
The cases where the focal player in Q-learning mode are with 0, 1, 2, 3, 4 TFT players in the neighborhood are computed separately in (a)-(e). 
 The black dashed line in both plots corresponds to $\overline{\Delta Q}_{s_j}=0$ for reference.
Other parameters: $\varepsilon = 0.01$, $\alpha =0.1$, $\gamma =0.9$, $r=0.1$.
}
\label{fig:dQ_SM_TFT}
\end{figure*}

To examine the interaction between Q-learning and TFT players in details, we classify all Q-learning players into five categories according to the number of TFT players in their neighborhood (i.e. 0, 1, 2, 3, 4). Fig.~\ref{fig:dQ_SM_TFT} shows the evolution of cooperation preference in terms of  $\overline{\Delta Q}_{s_j}$ for Q-learners for all ten states in SM. When there are no TFT players in the Q-learning players, they prefer defection in any state [Fig.~\ref{fig:dQ_SM_TFT}(a)]. Fig.~\ref{fig:dQ_SM_TFT}(b) shows that Q-learning players start to choose cooperation when there are one TFT player in their neighborhood. Remarkably, when there are two or more TFT players in their neighborhood, Q-learning players always prefer cooperation within any state, shown Fig.~\ref{fig:dQ_SM_TFT}(c-e). This means that in Q-T interactions,  Q-learning players successfully learn to cooperate to maximize their rewards, and this cooperation propensity increases with the number of their TFT neighbors.

\bibliography{ref}

\begin{thebibliography}{66}%
\makeatletter
\providecommand \@ifxundefined [1]{%
 \@ifx{#1\undefined}
}%
\providecommand \@ifnum [1]{%
 \ifnum #1\expandafter \@firstoftwo
 \else \expandafter \@secondoftwo
 \fi
}%
\providecommand \@ifx [1]{%
 \ifx #1\expandafter \@firstoftwo
 \else \expandafter \@secondoftwo
 \fi
}%
\providecommand \natexlab [1]{#1}%
\providecommand \enquote  [1]{``#1''}%
\providecommand \bibnamefont  [1]{#1}%
\providecommand \bibfnamefont [1]{#1}%
\providecommand \citenamefont [1]{#1}%
\providecommand \href@noop [0]{\@secondoftwo}%
\providecommand \href [0]{\begingroup \@sanitize@url \@href}%
\providecommand \@href[1]{\@@startlink{#1}\@@href}%
\providecommand \@@href[1]{\endgroup#1\@@endlink}%
\providecommand \@sanitize@url [0]{\catcode `\\12\catcode `\$12\catcode
  `\&12\catcode `\#12\catcode `\^12\catcode `\_12\catcode `\%12\relax}%
\providecommand \@@startlink[1]{}%
\providecommand \@@endlink[0]{}%
\providecommand \url  [0]{\begingroup\@sanitize@url \@url }%
\providecommand \@url [1]{\endgroup\@href {#1}{\urlprefix }}%
\providecommand \urlprefix  [0]{URL }%
\providecommand \Eprint [0]{\href }%
\providecommand \doibase [0]{http://dx.doi.org/}%
\providecommand \selectlanguage [0]{\@gobble}%
\providecommand \bibinfo  [0]{\@secondoftwo}%
\providecommand \bibfield  [0]{\@secondoftwo}%
\providecommand \translation [1]{[#1]}%
\providecommand \BibitemOpen [0]{}%
\providecommand \bibitemStop [0]{}%
\providecommand \bibitemNoStop [0]{.\EOS\space}%
\providecommand \EOS [0]{\spacefactor3000\relax}%
\providecommand \BibitemShut  [1]{\csname bibitem#1\endcsname}%
\let\auto@bib@innerbib\@empty
\bibitem [{\citenamefont {Maynard~Smith}\ and\ \citenamefont
  {Szathmary}(1997)}]{smith1997major}%
  \BibitemOpen
  \bibfield  {author} {\bibinfo {author} {\bibfnamefont {J.}~\bibnamefont
  {Maynard~Smith}}\ and\ \bibinfo {author} {\bibfnamefont {E.}~\bibnamefont
  {Szathmary}},\ }\href@noop {} {\emph {\bibinfo {title} {The Major Transitions
  in Evolution}}}\ (\bibinfo  {publisher} {OUP Oxford},\ \bibinfo {year}
  {1997})\BibitemShut {NoStop}%
\bibitem [{\citenamefont {Zelenski}\ \emph {et~al.}(2015)\citenamefont
  {Zelenski}, \citenamefont {Dopko},\ and\ \citenamefont
  {Capaldi}}]{Zelenski2015CooperationII}%
  \BibitemOpen
  \bibfield  {author} {\bibinfo {author} {\bibfnamefont {J.~M.}\ \bibnamefont
  {Zelenski}}, \bibinfo {author} {\bibfnamefont {R.~L.}\ \bibnamefont {Dopko}},
  \ and\ \bibinfo {author} {\bibfnamefont {C.~A.}\ \bibnamefont {Capaldi}},\
  }\href {https://api.semanticscholar.org/CorpusID:49536672} {\bibfield
  {journal} {\bibinfo  {journal} {Journal of Environmental Psychology}\
  }\textbf {\bibinfo {volume} {42}},\ \bibinfo {pages} {24} (\bibinfo {year}
  {2015})}\BibitemShut {NoStop}%
\bibitem [{\citenamefont {Cheney}(2011)}]{Cheney2011Extent}%
  \BibitemOpen
  \bibfield  {author} {\bibinfo {author} {\bibfnamefont {D.~L.}\ \bibnamefont
  {Cheney}},\ }\href {\doibase 10.1073/pnas.1100291108} {\bibfield  {journal}
  {\bibinfo  {journal} {Proceedings of the National Academy of Sciences}\
  }\textbf {\bibinfo {volume} {108}},\ \bibinfo {pages} {10902} (\bibinfo
  {year} {2011})}\BibitemShut {NoStop}%
\bibitem [{\citenamefont {Dawkins}(2006)}]{dawkins2016selfish}%
  \BibitemOpen
  \bibfield  {author} {\bibinfo {author} {\bibfnamefont {R.}~\bibnamefont
  {Dawkins}},\ }\href@noop {} {\emph {\bibinfo {title} {The Selfish Gene}}}\
  (\bibinfo  {publisher} {Oxford University Press},\ \bibinfo {address} {New
  York, US},\ \bibinfo {year} {2006})\BibitemShut {NoStop}%
\bibitem [{\citenamefont {Rapoport}\ and\ \citenamefont
  {Chammah}(1965)}]{rapoport1965prisoner}%
  \BibitemOpen
  \bibfield  {author} {\bibinfo {author} {\bibfnamefont {A.}~\bibnamefont
  {Rapoport}}\ and\ \bibinfo {author} {\bibfnamefont {A.~M.}\ \bibnamefont
  {Chammah}},\ }\href@noop {} {\emph {\bibinfo {title} {Prisoner's dilemma: A
  study in conflict and cooperation}}},\ Vol.\ \bibinfo {volume} {165}\
  (\bibinfo  {publisher} {University of Michigan press},\ \bibinfo {year}
  {1965})\BibitemShut {NoStop}%
\bibitem [{\citenamefont {Pennisi}(2005)}]{Elizabeth2005How}%
  \BibitemOpen
  \bibfield  {author} {\bibinfo {author} {\bibfnamefont {E.}~\bibnamefont
  {Pennisi}},\ }\href {\doibase 10.1126/science.309.5731.93} {\bibfield
  {journal} {\bibinfo  {journal} {Science}\ }\textbf {\bibinfo {volume}
  {309}},\ \bibinfo {pages} {93} (\bibinfo {year} {2005})}\BibitemShut
  {NoStop}%
\bibitem [{\citenamefont {Hardin}(1968)}]{Garrett1968The}%
  \BibitemOpen
  \bibfield  {author} {\bibinfo {author} {\bibfnamefont {G.}~\bibnamefont
  {Hardin}},\ }\href {\doibase 10.1126/science.162.3859.1243} {\bibfield
  {journal} {\bibinfo  {journal} {Science}\ }\textbf {\bibinfo {volume}
  {162}},\ \bibinfo {pages} {1243} (\bibinfo {year} {1968})}\BibitemShut
  {NoStop}%
\bibitem [{\citenamefont {Kollock}(1998)}]{Kollock1998Social}%
  \BibitemOpen
  \bibfield  {author} {\bibinfo {author} {\bibfnamefont {P.}~\bibnamefont
  {Kollock}},\ }\href {\doibase 10.1146/annurev.soc.24.1.183} {\bibfield
  {journal} {\bibinfo  {journal} {Annual Review of Sociology}\ }\textbf
  {\bibinfo {volume} {24}},\ \bibinfo {pages} {183} (\bibinfo {year}
  {1998})}\BibitemShut {NoStop}%
\bibitem [{\citenamefont {Milinski}\ \emph {et~al.}(2006)\citenamefont
  {Milinski}, \citenamefont {Semmann}, \citenamefont {Krambeck},\ and\
  \citenamefont {Marotzke}}]{Manfred2006Stabilizing}%
  \BibitemOpen
  \bibfield  {author} {\bibinfo {author} {\bibfnamefont {M.}~\bibnamefont
  {Milinski}}, \bibinfo {author} {\bibfnamefont {D.}~\bibnamefont {Semmann}},
  \bibinfo {author} {\bibfnamefont {H.-J.}\ \bibnamefont {Krambeck}}, \ and\
  \bibinfo {author} {\bibfnamefont {J.}~\bibnamefont {Marotzke}},\ }\href
  {\doibase 10.1073/pnas.0504902103} {\bibfield  {journal} {\bibinfo  {journal}
  {Proceedings of the National Academy of Sciences}\ }\textbf {\bibinfo
  {volume} {103}},\ \bibinfo {pages} {3994} (\bibinfo {year}
  {2006})}\BibitemShut {NoStop}%
\bibitem [{\citenamefont {Poundstone}(1993)}]{poundstone1993prisoner}%
  \BibitemOpen
  \bibfield  {author} {\bibinfo {author} {\bibfnamefont {W.}~\bibnamefont
  {Poundstone}},\ }\href@noop {} {\emph {\bibinfo {title} {Prisoner's dilemma:
  John von Neumann, game theory, and the puzzle of the bomb}}}\ (\bibinfo
  {publisher} {Anchor},\ \bibinfo {year} {1993})\BibitemShut {NoStop}%
\bibitem [{\citenamefont {Doebeli}\ and\ \citenamefont
  {Hauert}()}]{doebeli2005models}%
  \BibitemOpen
  \bibfield  {author} {\bibinfo {author} {\bibfnamefont {M.}~\bibnamefont
  {Doebeli}}\ and\ \bibinfo {author} {\bibfnamefont {C.}~\bibnamefont
  {Hauert}},\ }\href {\doibase
  https://doi.org/10.1111/j.1461-0248.2005.00773.x} {\bibfield  {journal}
  {\bibinfo  {journal} {Ecology Letters}\ }\textbf {\bibinfo {volume} {8}},\
  \bibinfo {pages} {748}}\BibitemShut {NoStop}%
\bibitem [{\citenamefont {Axelrod}(1980{\natexlab{a}})}]{axelrod1980effective}%
  \BibitemOpen
  \bibfield  {author} {\bibinfo {author} {\bibfnamefont {R.}~\bibnamefont
  {Axelrod}},\ }\href {\doibase 10.1177/002200278002400101} {\bibfield
  {journal} {\bibinfo  {journal} {Journal of Conflict Resolution}\ }\textbf
  {\bibinfo {volume} {24}},\ \bibinfo {pages} {3} (\bibinfo {year}
  {1980}{\natexlab{a}})}\BibitemShut {NoStop}%
\bibitem [{\citenamefont {Perc}\ and\ \citenamefont
  {Szolnoki}(2008)}]{perc2008social}%
  \BibitemOpen
  \bibfield  {author} {\bibinfo {author} {\bibfnamefont {M.}~\bibnamefont
  {Perc}}\ and\ \bibinfo {author} {\bibfnamefont {A.}~\bibnamefont
  {Szolnoki}},\ }\href {\doibase 10.1103/PhysRevE.77.011904} {\bibfield
  {journal} {\bibinfo  {journal} {Physical Review E}\ }\textbf {\bibinfo
  {volume} {77}},\ \bibinfo {pages} {011904} (\bibinfo {year}
  {2008})}\BibitemShut {NoStop}%
\bibitem [{\citenamefont {Axelrod}(1980{\natexlab{b}})}]{axelrod1980more}%
  \BibitemOpen
  \bibfield  {author} {\bibinfo {author} {\bibfnamefont {R.}~\bibnamefont
  {Axelrod}},\ }\href {\doibase 10.1177/002200278002400301} {\bibfield
  {journal} {\bibinfo  {journal} {Journal of Conflict Resolution}\ }\textbf
  {\bibinfo {volume} {24}},\ \bibinfo {pages} {379} (\bibinfo {year}
  {1980}{\natexlab{b}})}\BibitemShut {NoStop}%
\bibitem [{\citenamefont {Milinski}\ and\ \citenamefont
  {Wedekind}(1998)}]{milinski1998working}%
  \BibitemOpen
  \bibfield  {author} {\bibinfo {author} {\bibfnamefont {M.}~\bibnamefont
  {Milinski}}\ and\ \bibinfo {author} {\bibfnamefont {C.}~\bibnamefont
  {Wedekind}},\ }\href {\doibase 10.1073/pnas.95.23.13755} {\bibfield
  {journal} {\bibinfo  {journal} {Proceedings of the National Academy of
  Sciences}\ }\textbf {\bibinfo {volume} {95}},\ \bibinfo {pages} {13755}
  (\bibinfo {year} {1998})}\BibitemShut {NoStop}%
\bibitem [{\citenamefont {Nowak}(2006)}]{Nowak2006Five}%
  \BibitemOpen
  \bibfield  {author} {\bibinfo {author} {\bibfnamefont {M.~A.}\ \bibnamefont
  {Nowak}},\ }\href {\doibase 10.1126/science.1133755} {\bibfield  {journal}
  {\bibinfo  {journal} {Science}\ }\textbf {\bibinfo {volume} {314}},\ \bibinfo
  {pages} {1560} (\bibinfo {year} {2006})}\BibitemShut {NoStop}%
\bibitem [{\citenamefont {Perc}\ \emph {et~al.}(2017)\citenamefont {Perc},
  \citenamefont {Jordan}, \citenamefont {Rand}, \citenamefont {Wang},
  \citenamefont {Boccaletti},\ and\ \citenamefont
  {Szolnoki}}]{Perc2017statistical}%
  \BibitemOpen
  \bibfield  {author} {\bibinfo {author} {\bibfnamefont {M.}~\bibnamefont
  {Perc}}, \bibinfo {author} {\bibfnamefont {J.~J.}\ \bibnamefont {Jordan}},
  \bibinfo {author} {\bibfnamefont {D.~G.}\ \bibnamefont {Rand}}, \bibinfo
  {author} {\bibfnamefont {Z.}~\bibnamefont {Wang}}, \bibinfo {author}
  {\bibfnamefont {S.}~\bibnamefont {Boccaletti}}, \ and\ \bibinfo {author}
  {\bibfnamefont {A.}~\bibnamefont {Szolnoki}},\ }\href {\doibase
  https://doi.org/10.1016/j.physrep.2017.05.004} {\bibfield  {journal}
  {\bibinfo  {journal} {Physics Reports}\ }\textbf {\bibinfo {volume} {687}},\
  \bibinfo {pages} {1} (\bibinfo {year} {2017})}\BibitemShut {NoStop}%
\bibitem [{\citenamefont {Hamilton}(1964)}]{Hamilton1964the}%
  \BibitemOpen
  \bibfield  {author} {\bibinfo {author} {\bibfnamefont {W.}~\bibnamefont
  {Hamilton}},\ }\href {\doibase https://doi.org/10.1016/0022-5193(64)90039-6}
  {\bibfield  {journal} {\bibinfo  {journal} {Journal of Theoretical Biology}\
  }\textbf {\bibinfo {volume} {7}},\ \bibinfo {pages} {17} (\bibinfo {year}
  {1964})}\BibitemShut {NoStop}%
\bibitem [{\citenamefont {Trivers}(1971)}]{trivers1971evolution}%
  \BibitemOpen
  \bibfield  {author} {\bibinfo {author} {\bibfnamefont {R.~L.}\ \bibnamefont
  {Trivers}},\ }\href@noop {} {\bibfield  {journal} {\bibinfo  {journal} {The
  Quarterly review of biology}\ }\textbf {\bibinfo {volume} {46}},\ \bibinfo
  {pages} {35} (\bibinfo {year} {1971})}\BibitemShut {NoStop}%
\bibitem [{\citenamefont {Nowak}\ and\ \citenamefont
  {Sigmund}(1998)}]{nowak1998evolution}%
  \BibitemOpen
  \bibfield  {author} {\bibinfo {author} {\bibfnamefont {M.~A.}\ \bibnamefont
  {Nowak}}\ and\ \bibinfo {author} {\bibfnamefont {K.}~\bibnamefont
  {Sigmund}},\ }\href {\doibase https://doi.org/10.1038/31225} {\bibfield
  {journal} {\bibinfo  {journal} {Nature}\ }\textbf {\bibinfo {volume} {393}},\
  \bibinfo {pages} {573} (\bibinfo {year} {1998})}\BibitemShut {NoStop}%
\bibitem [{\citenamefont {Nowak}\ and\ \citenamefont
  {May}(1992)}]{nowak1992evolutionary}%
  \BibitemOpen
  \bibfield  {author} {\bibinfo {author} {\bibfnamefont {M.~A.}\ \bibnamefont
  {Nowak}}\ and\ \bibinfo {author} {\bibfnamefont {R.~M.}\ \bibnamefont
  {May}},\ }\href {\doibase https://doi.org/10.1038/359826a0} {\bibfield
  {journal} {\bibinfo  {journal} {Nature}\ }\textbf {\bibinfo {volume} {359}},\
  \bibinfo {pages} {826} (\bibinfo {year} {1992})}\BibitemShut {NoStop}%
\bibitem [{\citenamefont {Wang}\ \emph {et~al.}(2013)\citenamefont {Wang},
  \citenamefont {Szolnoki},\ and\ \citenamefont
  {Perc}}]{Wang2013Interdependent}%
  \BibitemOpen
  \bibfield  {author} {\bibinfo {author} {\bibfnamefont {Z.}~\bibnamefont
  {Wang}}, \bibinfo {author} {\bibfnamefont {A.}~\bibnamefont {Szolnoki}}, \
  and\ \bibinfo {author} {\bibfnamefont {M.}~\bibnamefont {Perc}},\ }\href
  {\doibase 10.1038/srep01183} {\bibfield  {journal} {\bibinfo  {journal}
  {Scientific Reports}\ }\textbf {\bibinfo {volume} {3}},\ \bibinfo {pages}
  {1183} (\bibinfo {year} {2013})}\BibitemShut {NoStop}%
\bibitem [{\citenamefont {Liang}\ \emph {et~al.}(2022)\citenamefont {Liang},
  \citenamefont {Wang}, \citenamefont {Zhang}, \citenamefont {Zheng},
  \citenamefont {Ma},\ and\ \citenamefont {Chen}}]{Liang2022dynamical}%
  \BibitemOpen
  \bibfield  {author} {\bibinfo {author} {\bibfnamefont {R.}~\bibnamefont
  {Liang}}, \bibinfo {author} {\bibfnamefont {Q.}~\bibnamefont {Wang}},
  \bibinfo {author} {\bibfnamefont {J.}~\bibnamefont {Zhang}}, \bibinfo
  {author} {\bibfnamefont {G.}~\bibnamefont {Zheng}}, \bibinfo {author}
  {\bibfnamefont {L.}~\bibnamefont {Ma}}, \ and\ \bibinfo {author}
  {\bibfnamefont {L.}~\bibnamefont {Chen}},\ }\href {\doibase
  https://doi.org/10.1103/PhysRevE.105.054302} {\bibfield  {journal} {\bibinfo
  {journal} {Physical Review E}\ }\textbf {\bibinfo {volume} {105}},\ \bibinfo
  {pages} {054302} (\bibinfo {year} {2022})}\BibitemShut {NoStop}%
\bibitem [{\citenamefont {Keller}(1999)}]{keller1999levels}%
  \BibitemOpen
  \bibfield  {author} {\bibinfo {author} {\bibfnamefont {L.}~\bibnamefont
  {Keller}},\ }\href@noop {} {\emph {\bibinfo {title} {Levels of selection in
  evolution}}},\ Vol.~\bibinfo {volume} {66}\ (\bibinfo  {publisher} {Princeton
  University Press},\ \bibinfo {year} {1999})\BibitemShut {NoStop}%
\bibitem [{\citenamefont {Maynard~Smith}(1964)}]{smith1964group}%
  \BibitemOpen
  \bibfield  {author} {\bibinfo {author} {\bibfnamefont {J.}~\bibnamefont
  {Maynard~Smith}},\ }\href {\doibase https://doi.org/10.1038/2011145a0}
  {\bibfield  {journal} {\bibinfo  {journal} {Nature}\ }\textbf {\bibinfo
  {volume} {201}},\ \bibinfo {pages} {1145} (\bibinfo {year}
  {1964})}\BibitemShut {NoStop}%
\bibitem [{\citenamefont {Sigmund}\ \emph {et~al.}(2001)\citenamefont
  {Sigmund}, \citenamefont {Hauert},\ and\ \citenamefont
  {Nowak}}]{Sigmund2001Reward}%
  \BibitemOpen
  \bibfield  {author} {\bibinfo {author} {\bibfnamefont {K.}~\bibnamefont
  {Sigmund}}, \bibinfo {author} {\bibfnamefont {C.}~\bibnamefont {Hauert}}, \
  and\ \bibinfo {author} {\bibfnamefont {M.~A.}\ \bibnamefont {Nowak}},\ }\href
  {\doibase 10.1073/pnas.161155698} {\bibfield  {journal} {\bibinfo  {journal}
  {Proceedings of the National Academy of Sciences}\ }\textbf {\bibinfo
  {volume} {98}},\ \bibinfo {pages} {10757} (\bibinfo {year}
  {2001})}\BibitemShut {NoStop}%
\bibitem [{\citenamefont {Xia}\ \emph {et~al.}(2023)\citenamefont {Xia},
  \citenamefont {Wang}, \citenamefont {Perc},\ and\ \citenamefont
  {Wang}}]{Xia2023Reputation}%
  \BibitemOpen
  \bibfield  {author} {\bibinfo {author} {\bibfnamefont {C.}~\bibnamefont
  {Xia}}, \bibinfo {author} {\bibfnamefont {J.}~\bibnamefont {Wang}}, \bibinfo
  {author} {\bibfnamefont {M.}~\bibnamefont {Perc}}, \ and\ \bibinfo {author}
  {\bibfnamefont {Z.}~\bibnamefont {Wang}},\ }\href {\doibase
  https://doi.org/10.1016/j.plrev.2023.05.002} {\bibfield  {journal} {\bibinfo
  {journal} {Physics of Life Reviews}\ }\textbf {\bibinfo {volume} {46}},\
  \bibinfo {pages} {8} (\bibinfo {year} {2023})}\BibitemShut {NoStop}%
\bibitem [{\citenamefont {Santos}\ \emph {et~al.}(2008)\citenamefont {Santos},
  \citenamefont {Santos},\ and\ \citenamefont {Pacheco}}]{Santos2008Social}%
  \BibitemOpen
  \bibfield  {author} {\bibinfo {author} {\bibfnamefont {F.~C.}\ \bibnamefont
  {Santos}}, \bibinfo {author} {\bibfnamefont {M.~D.}\ \bibnamefont {Santos}},
  \ and\ \bibinfo {author} {\bibfnamefont {J.~M.}\ \bibnamefont {Pacheco}},\
  }\href {\doibase 10.1038/nature06940} {\bibfield  {journal} {\bibinfo
  {journal} {Nature}\ }\textbf {\bibinfo {volume} {454}},\ \bibinfo {pages}
  {213} (\bibinfo {year} {2008})}\BibitemShut {NoStop}%
\bibitem [{\citenamefont {Liang}\ \emph {et~al.}(2021)\citenamefont {Liang},
  \citenamefont {Zhang}, \citenamefont {Zheng},\ and\ \citenamefont
  {Chen}}]{Liang2021Social}%
  \BibitemOpen
  \bibfield  {author} {\bibinfo {author} {\bibfnamefont {R.}~\bibnamefont
  {Liang}}, \bibinfo {author} {\bibfnamefont {J.}~\bibnamefont {Zhang}},
  \bibinfo {author} {\bibfnamefont {G.}~\bibnamefont {Zheng}}, \ and\ \bibinfo
  {author} {\bibfnamefont {L.}~\bibnamefont {Chen}},\ }\href {\doibase
  https://doi.org/10.1016/j.physa.2020.125726} {\bibfield  {journal} {\bibinfo
  {journal} {Physica A}\ }\textbf {\bibinfo {volume} {567}},\ \bibinfo {pages}
  {125726} (\bibinfo {year} {2021})}\BibitemShut {NoStop}%
\bibitem [{\citenamefont {Watkins}\ and\ \citenamefont
  {Dayan}(1992{\natexlab{a}})}]{Watkins1992Technical}%
  \BibitemOpen
  \bibfield  {author} {\bibinfo {author} {\bibfnamefont {C.}~\bibnamefont
  {Watkins}}\ and\ \bibinfo {author} {\bibfnamefont {P.}~\bibnamefont
  {Dayan}},\ }\href@noop {} {\bibfield  {journal} {\bibinfo  {journal} {Machine
  Learning}\ }\textbf {\bibinfo {volume} {8}},\ \bibinfo {pages} {279}
  (\bibinfo {year} {1992}{\natexlab{a}})}\BibitemShut {NoStop}%
\bibitem [{\citenamefont {Sutton}\ and\ \citenamefont
  {Barto}(2018)}]{Sutton2018reinforcement}%
  \BibitemOpen
  \bibfield  {author} {\bibinfo {author} {\bibfnamefont {R.~S.}\ \bibnamefont
  {Sutton}}\ and\ \bibinfo {author} {\bibfnamefont {A.~G.}\ \bibnamefont
  {Barto}},\ }\href@noop {} {\emph {\bibinfo {title} {Reinforcement Learning:
  An Introduction}}}\ (\bibinfo  {publisher} {MIT press},\ \bibinfo {year}
  {2018})\BibitemShut {NoStop}%
\bibitem [{\citenamefont {Bandura}\ and\ \citenamefont
  {Walters}(1977)}]{Bandura1977social}%
  \BibitemOpen
  \bibfield  {author} {\bibinfo {author} {\bibfnamefont {A.}~\bibnamefont
  {Bandura}}\ and\ \bibinfo {author} {\bibfnamefont {R.~H.}\ \bibnamefont
  {Walters}},\ }\href@noop {} {\emph {\bibinfo {title} {Social Learning
  Theory}}}\ (\bibinfo  {publisher} {Englewood cliffs Prentice Hall},\ \bibinfo
  {year} {1977})\BibitemShut {NoStop}%
\bibitem [{\citenamefont {Silver}\ \emph {et~al.}(2018)\citenamefont {Silver},
  \citenamefont {Hubert}, \citenamefont {Schrittwieser}, \citenamefont
  {Antonoglou}, \citenamefont {Lai}, \citenamefont {Guez}, \citenamefont
  {Lanctot}, \citenamefont {Sifre}, \citenamefont {Kumaran}, \citenamefont
  {Graepel}, \citenamefont {Lillicrap}, \citenamefont {Simonyan},\ and\
  \citenamefont {Hassabis}}]{silver2018general}%
  \BibitemOpen
  \bibfield  {author} {\bibinfo {author} {\bibfnamefont {D.}~\bibnamefont
  {Silver}}, \bibinfo {author} {\bibfnamefont {T.}~\bibnamefont {Hubert}},
  \bibinfo {author} {\bibfnamefont {J.}~\bibnamefont {Schrittwieser}}, \bibinfo
  {author} {\bibfnamefont {I.}~\bibnamefont {Antonoglou}}, \bibinfo {author}
  {\bibfnamefont {M.}~\bibnamefont {Lai}}, \bibinfo {author} {\bibfnamefont
  {A.}~\bibnamefont {Guez}}, \bibinfo {author} {\bibfnamefont {M.}~\bibnamefont
  {Lanctot}}, \bibinfo {author} {\bibfnamefont {L.}~\bibnamefont {Sifre}},
  \bibinfo {author} {\bibfnamefont {D.}~\bibnamefont {Kumaran}}, \bibinfo
  {author} {\bibfnamefont {T.}~\bibnamefont {Graepel}}, \bibinfo {author}
  {\bibfnamefont {T.}~\bibnamefont {Lillicrap}}, \bibinfo {author}
  {\bibfnamefont {K.}~\bibnamefont {Simonyan}}, \ and\ \bibinfo {author}
  {\bibfnamefont {D.}~\bibnamefont {Hassabis}},\ }\href {\doibase
  10.1126/science.aar6404} {\bibfield  {journal} {\bibinfo  {journal}
  {Science}\ }\textbf {\bibinfo {volume} {362}},\ \bibinfo {pages} {1140}
  (\bibinfo {year} {2018})}\BibitemShut {NoStop}%
\bibitem [{\citenamefont {Lee}\ \emph {et~al.}(2012)\citenamefont {Lee},
  \citenamefont {Seo},\ and\ \citenamefont {Jung}}]{Lee2012neural}%
  \BibitemOpen
  \bibfield  {author} {\bibinfo {author} {\bibfnamefont {D.}~\bibnamefont
  {Lee}}, \bibinfo {author} {\bibfnamefont {H.}~\bibnamefont {Seo}}, \ and\
  \bibinfo {author} {\bibfnamefont {M.~W.}\ \bibnamefont {Jung}},\ }\href
  {\doibase https://doi.org/10.1146/annurev-neuro-062111-150512} {\bibfield
  {journal} {\bibinfo  {journal} {Annual Review of Neuroscience}\ }\textbf
  {\bibinfo {volume} {35}},\ \bibinfo {pages} {287} (\bibinfo {year}
  {2012})}\BibitemShut {NoStop}%
\bibitem [{\citenamefont {Rangel}\ \emph {et~al.}(2008)\citenamefont {Rangel},
  \citenamefont {Camerer},\ and\ \citenamefont {Montague}}]{Rangel2008A}%
  \BibitemOpen
  \bibfield  {author} {\bibinfo {author} {\bibfnamefont {A.}~\bibnamefont
  {Rangel}}, \bibinfo {author} {\bibfnamefont {C.}~\bibnamefont {Camerer}}, \
  and\ \bibinfo {author} {\bibfnamefont {P.~R.}\ \bibnamefont {Montague}},\
  }\href {\doibase 10.1038/nrn2357} {\bibfield  {journal} {\bibinfo  {journal}
  {Nature Reviews Neuroscience}\ }\textbf {\bibinfo {volume} {9}},\ \bibinfo
  {pages} {545} (\bibinfo {year} {2008})}\BibitemShut {NoStop}%
\bibitem [{\citenamefont {Zhang}\ \emph {et~al.}(2020)\citenamefont {Zhang},
  \citenamefont {Zhang}, \citenamefont {Chen},\ and\ \citenamefont
  {Liu}}]{Zhang2020Oscillatory}%
  \BibitemOpen
  \bibfield  {author} {\bibinfo {author} {\bibfnamefont {S.}~\bibnamefont
  {Zhang}}, \bibinfo {author} {\bibfnamefont {J.}~\bibnamefont {Zhang}},
  \bibinfo {author} {\bibfnamefont {L.}~\bibnamefont {Chen}}, \ and\ \bibinfo
  {author} {\bibfnamefont {X.}~\bibnamefont {Liu}},\ }\href {\doibase
  10.1007/s11071-019-05398-4} {\bibfield  {journal} {\bibinfo  {journal}
  {Nonlinear Dynamics}\ }\textbf {\bibinfo {volume} {99}},\ \bibinfo {pages}
  {3301} (\bibinfo {year} {2020})}\BibitemShut {NoStop}%
\bibitem [{\citenamefont {Wang}\ \emph {et~al.}(2022)\citenamefont {Wang},
  \citenamefont {Jia}, \citenamefont {Zhang}, \citenamefont {Zhu},
  \citenamefont {Perc}, \citenamefont {Shi},\ and\ \citenamefont
  {Wang}}]{Wang2022Levy}%
  \BibitemOpen
  \bibfield  {author} {\bibinfo {author} {\bibfnamefont {L.}~\bibnamefont
  {Wang}}, \bibinfo {author} {\bibfnamefont {D.}~\bibnamefont {Jia}}, \bibinfo
  {author} {\bibfnamefont {L.}~\bibnamefont {Zhang}}, \bibinfo {author}
  {\bibfnamefont {P.}~\bibnamefont {Zhu}}, \bibinfo {author} {\bibfnamefont
  {M.}~\bibnamefont {Perc}}, \bibinfo {author} {\bibfnamefont {L.}~\bibnamefont
  {Shi}}, \ and\ \bibinfo {author} {\bibfnamefont {Z.}~\bibnamefont {Wang}},\
  }\href {\doibase 10.1007/s11071-022-07289-7} {\bibfield  {journal} {\bibinfo
  {journal} {Nonlinear Dynamics}\ }\textbf {\bibinfo {volume} {108}},\ \bibinfo
  {pages} {1837} (\bibinfo {year} {2022})}\BibitemShut {NoStop}%
\bibitem [{\citenamefont {Wang}\ \emph {et~al.}(2023)\citenamefont {Wang},
  \citenamefont {Fan}, \citenamefont {Zhang}, \citenamefont {Zou},\ and\
  \citenamefont {Wang}}]{Wang2023Synergistic}%
  \BibitemOpen
  \bibfield  {author} {\bibinfo {author} {\bibfnamefont {L.}~\bibnamefont
  {Wang}}, \bibinfo {author} {\bibfnamefont {L.}~\bibnamefont {Fan}}, \bibinfo
  {author} {\bibfnamefont {L.}~\bibnamefont {Zhang}}, \bibinfo {author}
  {\bibfnamefont {R.}~\bibnamefont {Zou}}, \ and\ \bibinfo {author}
  {\bibfnamefont {Z.}~\bibnamefont {Wang}},\ }\href {\doibase
  10.1088/1367-2630/acd26e} {\bibfield  {journal} {\bibinfo  {journal} {New
  Journal of Physics}\ }\textbf {\bibinfo {volume} {25}},\ \bibinfo {pages}
  {073008} (\bibinfo {year} {2023})}\BibitemShut {NoStop}%
\bibitem [{\citenamefont {He}\ \emph {et~al.}(2022)\citenamefont {He},
  \citenamefont {Geng}, \citenamefont {Du}, \citenamefont {Shi},\ and\
  \citenamefont {Wang}}]{He2022migration}%
  \BibitemOpen
  \bibfield  {author} {\bibinfo {author} {\bibfnamefont {Z.}~\bibnamefont
  {He}}, \bibinfo {author} {\bibfnamefont {Y.}~\bibnamefont {Geng}}, \bibinfo
  {author} {\bibfnamefont {C.}~\bibnamefont {Du}}, \bibinfo {author}
  {\bibfnamefont {L.}~\bibnamefont {Shi}}, \ and\ \bibinfo {author}
  {\bibfnamefont {Z.}~\bibnamefont {Wang}},\ }\href {\doibase
  10.1088/1367-2630/acadfd} {\bibfield  {journal} {\bibinfo  {journal} {New
  Journal of Physics}\ }\textbf {\bibinfo {volume} {24}},\ \bibinfo {pages}
  {123038} (\bibinfo {year} {2022})}\BibitemShut {NoStop}%
\bibitem [{\citenamefont {Ding}\ \emph {et~al.}(2023)\citenamefont {Ding},
  \citenamefont {Zheng}, \citenamefont {Cai}, \citenamefont {Cai},
  \citenamefont {Chen}, \citenamefont {Zhang},\ and\ \citenamefont
  {Wang}}]{Ding2023Emergence}%
  \BibitemOpen
  \bibfield  {author} {\bibinfo {author} {\bibfnamefont {Z.}~\bibnamefont
  {Ding}}, \bibinfo {author} {\bibfnamefont {G.}~\bibnamefont {Zheng}},
  \bibinfo {author} {\bibfnamefont {C.}~\bibnamefont {Cai}}, \bibinfo {author}
  {\bibfnamefont {W.}~\bibnamefont {Cai}}, \bibinfo {author} {\bibfnamefont
  {L.}~\bibnamefont {Chen}}, \bibinfo {author} {\bibfnamefont {J.}~\bibnamefont
  {Zhang}}, \ and\ \bibinfo {author} {\bibfnamefont {X.}~\bibnamefont {Wang}},\
  }\href {\doibase https://doi.org/10.1016/j.chaos.2023.114032} {\bibfield
  {journal} {\bibinfo  {journal} {Chaos, Solitons \& Fractals}\ }\textbf
  {\bibinfo {volume} {175}},\ \bibinfo {pages} {114032} (\bibinfo {year}
  {2023})}\BibitemShut {NoStop}%
\bibitem [{\citenamefont {Geng}\ \emph {et~al.}(2022)\citenamefont {Geng},
  \citenamefont {Liu}, \citenamefont {Lu}, \citenamefont {Shen},\ and\
  \citenamefont {Shi}}]{Geng2022Reinforcement}%
  \BibitemOpen
  \bibfield  {author} {\bibinfo {author} {\bibfnamefont {Y.}~\bibnamefont
  {Geng}}, \bibinfo {author} {\bibfnamefont {Y.}~\bibnamefont {Liu}}, \bibinfo
  {author} {\bibfnamefont {Y.}~\bibnamefont {Lu}}, \bibinfo {author}
  {\bibfnamefont {C.}~\bibnamefont {Shen}}, \ and\ \bibinfo {author}
  {\bibfnamefont {L.}~\bibnamefont {Shi}},\ }\href {\doibase
  https://doi.org/10.1016/j.amc.2022.127182} {\bibfield  {journal} {\bibinfo
  {journal} {Applied Mathematics and Computation}\ }\textbf {\bibinfo {volume}
  {427}},\ \bibinfo {pages} {127182} (\bibinfo {year} {2022})}\BibitemShut
  {NoStop}%
\bibitem [{\citenamefont {Zhang}\ \emph
  {et~al.}(2024{\natexlab{a}})\citenamefont {Zhang}, \citenamefont {Rong},
  \citenamefont {Zheng}, \citenamefont {Zhang},\ and\ \citenamefont
  {Chen}}]{Zhang2024emergence}%
  \BibitemOpen
  \bibfield  {author} {\bibinfo {author} {\bibfnamefont {J.}~\bibnamefont
  {Zhang}}, \bibinfo {author} {\bibfnamefont {Z.}~\bibnamefont {Rong}},
  \bibinfo {author} {\bibfnamefont {G.}~\bibnamefont {Zheng}}, \bibinfo
  {author} {\bibfnamefont {J.}~\bibnamefont {Zhang}}, \ and\ \bibinfo {author}
  {\bibfnamefont {L.}~\bibnamefont {Chen}},\ }\href {\doibase
  10.1088/2632-072X/ad3f65} {\bibfield  {journal} {\bibinfo  {journal} {Journal
  of Physics: Complexity}\ }\textbf {\bibinfo {volume} {5}},\ \bibinfo {pages}
  {025006} (\bibinfo {year} {2024}{\natexlab{a}})}\BibitemShut {NoStop}%
\bibitem [{\citenamefont {Zheng}\ \emph {et~al.}(2024)\citenamefont {Zheng},
  \citenamefont {Zhang}, \citenamefont {Zhang}, \citenamefont {Cai},\ and\
  \citenamefont {Chen}}]{zheng2023decoding}%
  \BibitemOpen
  \bibfield  {author} {\bibinfo {author} {\bibfnamefont {G.}~\bibnamefont
  {Zheng}}, \bibinfo {author} {\bibfnamefont {J.}~\bibnamefont {Zhang}},
  \bibinfo {author} {\bibfnamefont {J.}~\bibnamefont {Zhang}}, \bibinfo
  {author} {\bibfnamefont {W.}~\bibnamefont {Cai}}, \ and\ \bibinfo {author}
  {\bibfnamefont {L.}~\bibnamefont {Chen}},\ }\href {\doibase
  https://doi.org/10.1088/1367-2630/ad4b5a} {\bibfield  {journal} {\bibinfo
  {journal} {New Journal of Physics}\ }\textbf {\bibinfo {volume} {26}},\
  \bibinfo {pages} {053041} (\bibinfo {year} {2024})}\BibitemShut {NoStop}%
\bibitem [{\citenamefont {Andrecut}\ and\ \citenamefont
  {Ali}(2001)}]{Andrecut2001q}%
  \BibitemOpen
  \bibfield  {author} {\bibinfo {author} {\bibfnamefont {M.}~\bibnamefont
  {Andrecut}}\ and\ \bibinfo {author} {\bibfnamefont {M.}~\bibnamefont {Ali}},\
  }\href {\doibase https://doi.org/10.1103/PhysRevE.64.067103} {\bibfield
  {journal} {\bibinfo  {journal} {Physical Review E}\ }\textbf {\bibinfo
  {volume} {64}},\ \bibinfo {pages} {067103} (\bibinfo {year}
  {2001})}\BibitemShut {NoStop}%
\bibitem [{\citenamefont {Zhang}\ \emph {et~al.}(2019)\citenamefont {Zhang},
  \citenamefont {Dong}, \citenamefont {Liu}, \citenamefont {Huang},
  \citenamefont {Huang},\ and\ \citenamefont {Lai}}]{Zhang2019reinforcement}%
  \BibitemOpen
  \bibfield  {author} {\bibinfo {author} {\bibfnamefont {S.}~\bibnamefont
  {Zhang}}, \bibinfo {author} {\bibfnamefont {J.}~\bibnamefont {Dong}},
  \bibinfo {author} {\bibfnamefont {L.}~\bibnamefont {Liu}}, \bibinfo {author}
  {\bibfnamefont {Z.}~\bibnamefont {Huang}}, \bibinfo {author} {\bibfnamefont
  {L.}~\bibnamefont {Huang}}, \ and\ \bibinfo {author} {\bibfnamefont
  {Y.}~\bibnamefont {Lai}},\ }\href {\doibase
  https://doi.org/10.1103/PhysRevE.99.032302} {\bibfield  {journal} {\bibinfo
  {journal} {Physical Review E}\ }\textbf {\bibinfo {volume} {99}},\ \bibinfo
  {pages} {032302} (\bibinfo {year} {2019})}\BibitemShut {NoStop}%
\bibitem [{\citenamefont {Zheng}\ \emph {et~al.}(2023)\citenamefont {Zheng},
  \citenamefont {Cai}, \citenamefont {Qi}, \citenamefont {Zhang},\ and\
  \citenamefont {Chen}}]{zheng2023optimal}%
  \BibitemOpen
  \bibfield  {author} {\bibinfo {author} {\bibfnamefont {G.}~\bibnamefont
  {Zheng}}, \bibinfo {author} {\bibfnamefont {W.}~\bibnamefont {Cai}}, \bibinfo
  {author} {\bibfnamefont {G.}~\bibnamefont {Qi}}, \bibinfo {author}
  {\bibfnamefont {J.}~\bibnamefont {Zhang}}, \ and\ \bibinfo {author}
  {\bibfnamefont {L.}~\bibnamefont {Chen}},\ }\href@noop {} {\bibfield
  {journal} {\bibinfo  {journal} {arXiv preprint arXiv:2312.14970}\ } (\bibinfo
  {year} {2023})}\BibitemShut {NoStop}%
\bibitem [{\citenamefont {Zhang}\ \emph
  {et~al.}(2024{\natexlab{b}})\citenamefont {Zhang}, \citenamefont {Dong},
  \citenamefont {Zhang}, \citenamefont {L{\"u}}, \citenamefont {Wang},\ and\
  \citenamefont {Huang}}]{Zhang2024self}%
  \BibitemOpen
  \bibfield  {author} {\bibinfo {author} {\bibfnamefont {S.}~\bibnamefont
  {Zhang}}, \bibinfo {author} {\bibfnamefont {J.}~\bibnamefont {Dong}},
  \bibinfo {author} {\bibfnamefont {H.}~\bibnamefont {Zhang}}, \bibinfo
  {author} {\bibfnamefont {Y.}~\bibnamefont {L{\"u}}}, \bibinfo {author}
  {\bibfnamefont {J.}~\bibnamefont {Wang}}, \ and\ \bibinfo {author}
  {\bibfnamefont {Z.}~\bibnamefont {Huang}},\ }\href {\doibase
  10.1007/s11467-023-1378-z} {\bibfield  {journal} {\bibinfo  {journal}
  {Frontiers of Physics}\ }\textbf {\bibinfo {volume} {19}},\ \bibinfo {pages}
  {1} (\bibinfo {year} {2024}{\natexlab{b}})}\BibitemShut {NoStop}%
\bibitem [{\citenamefont {Tomov}\ \emph {et~al.}(2021)\citenamefont {Tomov},
  \citenamefont {Schulz},\ and\ \citenamefont {Gershman}}]{Tomov2021multi}%
  \BibitemOpen
  \bibfield  {author} {\bibinfo {author} {\bibfnamefont {M.~S.}\ \bibnamefont
  {Tomov}}, \bibinfo {author} {\bibfnamefont {E.}~\bibnamefont {Schulz}}, \
  and\ \bibinfo {author} {\bibfnamefont {S.~J.}\ \bibnamefont {Gershman}},\
  }\href {\doibase https://doi.org/10.1038/s41562-020-01035-y} {\bibfield
  {journal} {\bibinfo  {journal} {Nature Human Behaviour}\ }\textbf {\bibinfo
  {volume} {5}},\ \bibinfo {pages} {764} (\bibinfo {year} {2021})}\BibitemShut
  {NoStop}%
\bibitem [{\citenamefont {Shi}\ and\ \citenamefont
  {Rong}(2022)}]{Shi2022analysis}%
  \BibitemOpen
  \bibfield  {author} {\bibinfo {author} {\bibfnamefont {Y.}~\bibnamefont
  {Shi}}\ and\ \bibinfo {author} {\bibfnamefont {Z.}~\bibnamefont {Rong}},\
  }\href {\doibase https://doi.org/10.1109/TCSII.2022.3161655} {\bibfield
  {journal} {\bibinfo  {journal} {IEEE Transactions on Circuits and Systems II:
  Express Briefs}\ }\textbf {\bibinfo {volume} {69}},\ \bibinfo {pages} {2463}
  (\bibinfo {year} {2022})}\BibitemShut {NoStop}%
\bibitem [{\citenamefont {Ding}\ \emph {et~al.}(2024)\citenamefont {Ding},
  \citenamefont {Zhang}, \citenamefont {Zheng}, \citenamefont {Cai},
  \citenamefont {Cai}, \citenamefont {Chen},\ and\ \citenamefont
  {Wang}}]{Ding2024Emergence}%
  \BibitemOpen
  \bibfield  {author} {\bibinfo {author} {\bibfnamefont {Z.}~\bibnamefont
  {Ding}}, \bibinfo {author} {\bibfnamefont {J.}~\bibnamefont {Zhang}},
  \bibinfo {author} {\bibfnamefont {G.}~\bibnamefont {Zheng}}, \bibinfo
  {author} {\bibfnamefont {W.}~\bibnamefont {Cai}}, \bibinfo {author}
  {\bibfnamefont {C.}~\bibnamefont {Cai}}, \bibinfo {author} {\bibfnamefont
  {L.}~\bibnamefont {Chen}}, \ and\ \bibinfo {author} {\bibfnamefont
  {X.}~\bibnamefont {Wang}},\ }\href@noop {} {\bibfield  {journal} {\bibinfo
  {journal} {Chaos, Solitons \& Fractals}\ }\textbf {\bibinfo {volume} {184}},\
  \bibinfo {pages} {114971} (\bibinfo {year} {2024})}\BibitemShut {NoStop}%
\bibitem [{\citenamefont {Watkins}(1989)}]{Watkins1989learning}%
  \BibitemOpen
  \bibfield  {author} {\bibinfo {author} {\bibfnamefont {C.~J. C.~H.}\
  \bibnamefont {Watkins}},\ }\emph {\bibinfo {title} {Learning from delayed
  rewards (Ph.D. thesis)}},\ \href@noop {} {Ph.D. thesis} (\bibinfo {year}
  {1989})\BibitemShut {NoStop}%
\bibitem [{\citenamefont {Watkins}\ and\ \citenamefont
  {Dayan}(1992{\natexlab{b}})}]{watkins1992q}%
  \BibitemOpen
  \bibfield  {author} {\bibinfo {author} {\bibfnamefont {C.~J.}\ \bibnamefont
  {Watkins}}\ and\ \bibinfo {author} {\bibfnamefont {P.}~\bibnamefont
  {Dayan}},\ }\href {\doibase https://doi.org/10.1007/BF00992698} {\bibfield
  {journal} {\bibinfo  {journal} {Machine Learning}\ }\textbf {\bibinfo
  {volume} {8}},\ \bibinfo {pages} {279} (\bibinfo {year}
  {1992}{\natexlab{b}})}\BibitemShut {NoStop}%
\bibitem [{\citenamefont {Van~Hasselt}\ \emph {et~al.}(2016)\citenamefont
  {Van~Hasselt}, \citenamefont {Guez},\ and\ \citenamefont
  {Silver}}]{van2016deep}%
  \BibitemOpen
  \bibfield  {author} {\bibinfo {author} {\bibfnamefont {H.}~\bibnamefont
  {Van~Hasselt}}, \bibinfo {author} {\bibfnamefont {A.}~\bibnamefont {Guez}}, \
  and\ \bibinfo {author} {\bibfnamefont {D.}~\bibnamefont {Silver}},\ }in\
  \href {\doibase https://doi.org/10.1609/aaai.v30i1.10295} {\emph {\bibinfo
  {booktitle} {Proceedings of the AAAI conference on artificial
  intelligence}}},\ Vol.~\bibinfo {volume} {30}\ (\bibinfo {year}
  {2016})\BibitemShut {NoStop}%
\bibitem [{\citenamefont {Wunder}\ \emph {et~al.}(2010)\citenamefont {Wunder},
  \citenamefont {Littman},\ and\ \citenamefont {Babes}}]{wunder2010classes}%
  \BibitemOpen
  \bibfield  {author} {\bibinfo {author} {\bibfnamefont {M.}~\bibnamefont
  {Wunder}}, \bibinfo {author} {\bibfnamefont {M.}~\bibnamefont {Littman}}, \
  and\ \bibinfo {author} {\bibfnamefont {M.}~\bibnamefont {Babes}},\ }in\ \href
  {https://dl.acm.org/doi/abs/10.5555/3104322.3104470} {\emph {\bibinfo
  {booktitle} {Proceedings of the 27th International Conference on
  International Conference on Machine Learning}}},\ \bibinfo {series and
  number} {ICML'10}\ (\bibinfo  {publisher} {Omnipress},\ \bibinfo {address}
  {Madison, WI, USA},\ \bibinfo {year} {2010})\ p.\ \bibinfo {pages}
  {1167–1174}\BibitemShut {NoStop}%
\bibitem [{\citenamefont {Szab{\'o}}\ and\ \citenamefont
  {Fath}(2007)}]{Szabo2007evolutionary}%
  \BibitemOpen
  \bibfield  {author} {\bibinfo {author} {\bibfnamefont {G.}~\bibnamefont
  {Szab{\'o}}}\ and\ \bibinfo {author} {\bibfnamefont {G.}~\bibnamefont
  {Fath}},\ }\href {\doibase https://doi.org/10.1016/j.physrep.2007.04.004}
  {\bibfield  {journal} {\bibinfo  {journal} {Physics Reports}\ }\textbf
  {\bibinfo {volume} {446}},\ \bibinfo {pages} {97} (\bibinfo {year}
  {2007})}\BibitemShut {NoStop}%
\bibitem [{\citenamefont {Axelrod}\ and\ \citenamefont
  {Hamilton}(1981)}]{axelrod1981evolution}%
  \BibitemOpen
  \bibfield  {author} {\bibinfo {author} {\bibfnamefont {R.}~\bibnamefont
  {Axelrod}}\ and\ \bibinfo {author} {\bibfnamefont {W.~D.}\ \bibnamefont
  {Hamilton}},\ }\href {\doibase 10.1126/science.7466396} {\bibfield  {journal}
  {\bibinfo  {journal} {Science}\ }\textbf {\bibinfo {volume} {211}},\ \bibinfo
  {pages} {1390} (\bibinfo {year} {1981})}\BibitemShut {NoStop}%
\bibitem [{\citenamefont {Nowak}\ and\ \citenamefont
  {Sigmund}(1992)}]{nowak1992tit}%
  \BibitemOpen
  \bibfield  {author} {\bibinfo {author} {\bibfnamefont {M.~A.}\ \bibnamefont
  {Nowak}}\ and\ \bibinfo {author} {\bibfnamefont {K.}~\bibnamefont
  {Sigmund}},\ }\href {\doibase https://doi.org/10.1038/355250a0} {\bibfield
  {journal} {\bibinfo  {journal} {Nature}\ }\textbf {\bibinfo {volume} {355}},\
  \bibinfo {pages} {250} (\bibinfo {year} {1992})}\BibitemShut {NoStop}%
\bibitem [{\citenamefont {Nowak}\ and\ \citenamefont
  {Sigmund}(1993)}]{nowak1993strategy}%
  \BibitemOpen
  \bibfield  {author} {\bibinfo {author} {\bibfnamefont {M.}~\bibnamefont
  {Nowak}}\ and\ \bibinfo {author} {\bibfnamefont {K.}~\bibnamefont
  {Sigmund}},\ }\href {\doibase https://doi.org/10.1038/364056a0} {\bibfield
  {journal} {\bibinfo  {journal} {Nature}\ }\textbf {\bibinfo {volume} {364}},\
  \bibinfo {pages} {56} (\bibinfo {year} {1993})}\BibitemShut {NoStop}%
\bibitem [{\citenamefont {Traulsen}\ \emph {et~al.}(2010)\citenamefont
  {Traulsen}, \citenamefont {Semmann}, \citenamefont {Sommerfeld},
  \citenamefont {Krambeck},\ and\ \citenamefont
  {Milinski}}]{Traulsen2010Human}%
  \BibitemOpen
  \bibfield  {author} {\bibinfo {author} {\bibfnamefont {A.}~\bibnamefont
  {Traulsen}}, \bibinfo {author} {\bibfnamefont {D.}~\bibnamefont {Semmann}},
  \bibinfo {author} {\bibfnamefont {R.~D.}\ \bibnamefont {Sommerfeld}},
  \bibinfo {author} {\bibfnamefont {H.-J.}\ \bibnamefont {Krambeck}}, \ and\
  \bibinfo {author} {\bibfnamefont {M.}~\bibnamefont {Milinski}},\ }\href
  {\doibase 10.1073/pnas.0912515107} {\bibfield  {journal} {\bibinfo  {journal}
  {Proc. Natl. Acad. Sci. U.S.A.}\ }\textbf {\bibinfo {volume} {107}},\
  \bibinfo {pages} {2962} (\bibinfo {year} {2010})}\BibitemShut {NoStop}%
\bibitem [{\citenamefont {Szolnoki}\ and\ \citenamefont
  {Perc}(2015)}]{Szolnoki2015Conformity}%
  \BibitemOpen
  \bibfield  {author} {\bibinfo {author} {\bibfnamefont {A.}~\bibnamefont
  {Szolnoki}}\ and\ \bibinfo {author} {\bibfnamefont {M.}~\bibnamefont
  {Perc}},\ }\href {\doibase https://doi.org/10.1098/rsif.2014.1299} {\bibfield
   {journal} {\bibinfo  {journal} {Journal of The Royal Society Interface}\
  }\textbf {\bibinfo {volume} {12}},\ \bibinfo {pages} {20141299} (\bibinfo
  {year} {2015})}\BibitemShut {NoStop}%
\bibitem [{\citenamefont {Szolnoki}\ and\ \citenamefont
  {Chen}(2018)}]{Szolnoki2018Competition}%
  \BibitemOpen
  \bibfield  {author} {\bibinfo {author} {\bibfnamefont {A.}~\bibnamefont
  {Szolnoki}}\ and\ \bibinfo {author} {\bibfnamefont {X.}~\bibnamefont
  {Chen}},\ }\href {\doibase 10.1088/1367-2630/aade3c} {\bibfield  {journal}
  {\bibinfo  {journal} {New Journal of Physics}\ }\textbf {\bibinfo {volume}
  {20}},\ \bibinfo {pages} {093008} (\bibinfo {year} {2018})}\BibitemShut
  {NoStop}%
\bibitem [{\citenamefont {Amaral}\ and\ \citenamefont
  {Javarone}(2018)}]{Amaral2018Heterogeneous}%
  \BibitemOpen
  \bibfield  {author} {\bibinfo {author} {\bibfnamefont {M.~A.}\ \bibnamefont
  {Amaral}}\ and\ \bibinfo {author} {\bibfnamefont {M.~A.}\ \bibnamefont
  {Javarone}},\ }\href {\doibase 10.1103/PhysRevE.97.042305} {\bibfield
  {journal} {\bibinfo  {journal} {Phys. Rev. E}\ }\textbf {\bibinfo {volume}
  {97}},\ \bibinfo {pages} {042305} (\bibinfo {year} {2018})}\BibitemShut
  {NoStop}%
\bibitem [{\citenamefont {Masuda}(2012)}]{Masuda2012evolution}%
  \BibitemOpen
  \bibfield  {author} {\bibinfo {author} {\bibfnamefont {N.}~\bibnamefont
  {Masuda}},\ }\href {\doibase 10.1038/srep00646} {\bibfield  {journal}
  {\bibinfo  {journal} {Scientific Reports}\ }\textbf {\bibinfo {volume} {2}},\
  \bibinfo {pages} {1} (\bibinfo {year} {2012})}\BibitemShut {NoStop}%
\bibitem [{\citenamefont {Zheng}\ \emph {et~al.}(2022)\citenamefont {Zheng},
  \citenamefont {Zhang}, \citenamefont {Liang}, \citenamefont {Ma},\ and\
  \citenamefont {Chen}}]{Zheng2022probabilistic}%
  \BibitemOpen
  \bibfield  {author} {\bibinfo {author} {\bibfnamefont {G.}~\bibnamefont
  {Zheng}}, \bibinfo {author} {\bibfnamefont {J.}~\bibnamefont {Zhang}},
  \bibinfo {author} {\bibfnamefont {R.}~\bibnamefont {Liang}}, \bibinfo
  {author} {\bibfnamefont {L.}~\bibnamefont {Ma}}, \ and\ \bibinfo {author}
  {\bibfnamefont {L.}~\bibnamefont {Chen}},\ }\href {\doibase
  10.1088/2632-072X/ac86b3} {\bibfield  {journal} {\bibinfo  {journal} {Journal
  of Physics: Complexity}\ }\textbf {\bibinfo {volume} {3}},\ \bibinfo {pages}
  {035004} (\bibinfo {year} {2022})}\BibitemShut {NoStop}%
\bibitem [{\citenamefont {Ma}\ \emph {et~al.}(2023)\citenamefont {Ma},
  \citenamefont {Zhang}, \citenamefont {Zheng}, \citenamefont {Liang},\ and\
  \citenamefont {Chen}}]{ma2023emergence}%
  \BibitemOpen
  \bibfield  {author} {\bibinfo {author} {\bibfnamefont {L.}~\bibnamefont
  {Ma}}, \bibinfo {author} {\bibfnamefont {J.}~\bibnamefont {Zhang}}, \bibinfo
  {author} {\bibfnamefont {G.}~\bibnamefont {Zheng}}, \bibinfo {author}
  {\bibfnamefont {R.}~\bibnamefont {Liang}}, \ and\ \bibinfo {author}
  {\bibfnamefont {L.}~\bibnamefont {Chen}},\ }\href {\doibase
  https://doi.org/10.1016/j.chaos.2023.113452} {\bibfield  {journal} {\bibinfo
  {journal} {Chaos, Solitons \& Fractals}\ }\textbf {\bibinfo {volume} {171}},\
  \bibinfo {pages} {113452} (\bibinfo {year} {2023})}\BibitemShut {NoStop}%
\bibitem [{\citenamefont {Han}\ \emph {et~al.}(2022)\citenamefont {Han},
  \citenamefont {Zhao},\ and\ \citenamefont {Xia}}]{Han2022hybrid}%
  \BibitemOpen
  \bibfield  {author} {\bibinfo {author} {\bibfnamefont {X.}~\bibnamefont
  {Han}}, \bibinfo {author} {\bibfnamefont {X.}~\bibnamefont {Zhao}}, \ and\
  \bibinfo {author} {\bibfnamefont {H.}~\bibnamefont {Xia}},\ }\href {\doibase
  10.1016/j.chaos.2022.112684} {\bibfield  {journal} {\bibinfo  {journal}
  {Chaos, Solitons \& Fractals}\ }\textbf {\bibinfo {volume} {164}},\ \bibinfo
  {pages} {112684} (\bibinfo {year} {2022})}\BibitemShut {NoStop}%
\end{thebibliography}%
\end{document}